\documentclass[letterpaper, paper,11pt]{AAS}	%

\usepackage{bm}
\usepackage{amsmath}
\usepackage{amsmath}
\usepackage{amssymb}
\usepackage{subfigure}
\usepackage[colorlinks=true, pdfstartview=FitV, linkcolor=black, citecolor= black, urlcolor= black]{hyperref}
\usepackage{overcite}
\usepackage{footnpag}			      	%

\PaperNumber{XX-XXX}

\begin{document}

\title{P-Bifurcations in Stochastic Flutter Model Under Turbulence}

\author{
Sunia Tanweer\footnotemark[1] \footnotemark[2]
\ and Firas A. Khasawneh\footnotemark[2]
}
\footnotetext[1]{Dept. of Mechanical Engineering, Michigan State University, MI 48824, USA.}
\footnotetext[2]{Dept. of Computational Mathematics, Sciences and Engineering, Michigan State University, MI 48824, USA.}

\maketitle{}

\begin{abstract}
Aeroelastic flutter represents a critical nonlinear instability arising from the coupling between structural elasticity and unsteady aerodynamics. In deterministic settings, flutter onset is associated with bifurcations of invariant sets such as equilibria or limit cycles. However, under stochastic excitation, long-time system behavior is better described in terms of stationary probability distributions rather than trajectory-based attractors.
In this work, we present a topology-based framework to detect stochastic (P-)bifurcations in a two-degree-of-freedom aeroelastic system with structural nonlinearity. The method operates on high-dimensional stationary distributions reconstructed via kernel density estimation (KDE) and characterizes their structure using persistent homology. 
We compare bifurcation behavior across three excitation models: sinusoidal perturbations, Dryden turbulence, and von Kármán turbulence. While conventional time-domain and phase-space analyses reveal only modest differences between these models, the proposed homological bifurcation plots detect consistent shifts in bifurcation onset and topological structure. The approach enables automated and scalable analysis of stochastic bifurcations in complex dynamical systems.

\end{abstract}

\section{Introduction}

Aeroelastic systems manifest very complex phenomena due to interactions between aerodynamic forces and structural elements of flexible bodies. 
One of the most important phenomenon within these systems is flutter---a self-excited oscillation that may lead to structural failure due to large amplitude oscillations~\cite{Berci2021, Jones1996}. 

When the aerofoil is subject to random wind gust excitation or when uncertainties in the system paramaters are present, limit cycle oscillations (LCOs) arise due to structural nonlinearities and represent a key signature of instability. In other cases, the interaction between the aerofoil and unsteady fluid flows can lead to complex phenomena like intermittency that contribute to stall and flutter~\cite{Venkatramani2018}. As the angle of attack increases, the airflow may separate from the surface, causing oscillations that can be exacerbated by stochastic influences. The resulting flow patterns can significantly alter the stability of the aerofoil, making stochastic analysis crucial for predicting and mitigating flutter risks~\cite{Chen2022, Huang1987, Wu2022}.
The system behavior---inherently probabilistic under randomness---cannot be accurately analysed with classical trajectory-based notions of attractors. Instead, long-time dynamics are described by invariant probability distributions.

Stochastic bifurcation~\cite{SriNamachchivaya1990, Bass1999} is a critical concept in understanding the dynamics of aerofoil LCOs, stall and flutter~\cite{Venkatramani2018, Ketseas2024}, particularly under the influence of random disturbances. In the context of aerofoils, stochastic bifurcation helps identify the critical points where small changes in parameters can lead to large-scale shifts in behavior, such as transitioning from stable to unstable conditions~\cite{Bethi2018}. 

The interrelation between stochastic bifurcation and flutter remains a topic of active research. Recent studies have focused on the stochastic analysis of flutter in multi-stable aerofoils, where the role of random disturbances is central to understanding oscillatory behavior during stall conditions~\cite{Irani2016, Hao2018, Tripathi2022}. These investigations demonstrate that statistical methods can characterize bifurcation behavior and predict flutter conditions, which is vital for the design of safer and more efficient aerofoils~\cite{Hao2018}. However, most existing approaches rely heavily on visual inspection of low-dimensional phase portraits, time series, or probability density plots, which limits the number of state variables and parameters that can be examined simultaneously, such as here~\cite{Venkatramani2018}. Such low-dimensional projections can obscure essential features of the underlying dynamics in inherently high-dimensional aeroelastic systems. Moreover, the reliance on manual inspection precludes rigorous, automated analysis over large parameter ranges, making comprehensive uncertainty quantification and comparative studies difficult. In addition, prior stochastic flutter studies predominantly consider simplified sinusoidal or narrow-band excitation models, whereas modern aviation applications commonly employ physically grounded turbulence models such as the Dryden and von Kármán spectra~\cite{Beal1993}. In contrast, the present work provides a topology-based, automated framework for detecting stochastic flutter bifurcations directly from high-dimensional stationary distributions, and systematically compares bifurcation behavior across sinusoidal, Dryden, and von Kármán gust models.
 
In this work, we propose a topology-based approach that analyzes the full stationary distribution in the system’s four-dimensional phase space. 
{The primary contributions of this work are:}
\begin{itemize}
\item A topological data analysis (TDA)-based framework for detecting stochastic bifurcations directly from high-dimensional stationary distributions,
\item Demonstration that topological features provide sensitive indicators of bifurcation onset even when conventional analyses show minimal differences,
\item A systematic comparison of stochastic bifurcation behavior across excitation models with distinct temporal correlation structures.
\end{itemize}

\section{Mathematical Preliminaries}
\label{sec:maths}

\subsection{Stochastic Attractors and Invariant Distribution}

For deterministic dynamical systems, long-time behavior is characterized by invariant sets such as fixed points, limit cycles, or strange attractors. In contrast, stochastic dynamical systems do not admit trajectory-wise invariant sets due to continual random perturbations. Instead, their asymptotic behavior is described either in terms of {random attractors} or, equivalently under suitable conditions, invariant probability measures.

In this work, we adopt the framework of {pullback attractors} for random dynamical systems. A pullback attractor is a random, time-dependent set that attracts all initial conditions when trajectories are pulled back to the distant past and evolved forward under a fixed realization of the noise \cite{Arnold1998, Crauel1994}. When the stochastic system is ergodic, the pullback attractor induces a unique stationary (invariant) probability measure, which describes the long-time distribution of system states independently of initial conditions.

Rather than attempting to compute the pullback attractor itself, we will focus on the topology of the associated stationary probability distribution. Specifically, we reconstruct the invariant measure using long-time Monte Carlo sampling and kernel density estimation, and analyze the topology of its high-probability regions via superlevel-set persistent homology. Topological changes in this stationary distribution provide a robust and quantitative signature of phenomenological (P-)bifurcations in the stochastic system.

\subsection{Aerofoil Model}

A two-degree-of-freedom (2-DOF) aerofoil system, adopted from Venkatramani et al.~\cite{Venkatramani2018}, has been used to model the coupled pitch and plunge motions responsible for flutter. The aerofoil is free to oscillate vertically (plunge, \( \epsilon \)) and rotationally (pitch, \( \alpha \)) about its elastic axis. The governing equations of motion, expressed in nondimensional form, are written as

\begin{align}
\epsilon'' + x_{\alpha}\,\alpha'' 
+ 2\zeta_{\epsilon}\frac{\omega}{U}\,\epsilon'
+ \left(\frac{\omega}{U}\right)^{2}\!\big(\epsilon + \beta_{\epsilon}\epsilon^{3}\big) \nonumber\\
= -\frac{1}{\pi\,\mu}\,C_L(\tau), \\[4pt]
\frac{x_{\alpha}}{r_{\alpha}^{2}}\,\epsilon'' + \alpha''
+ 2\zeta_{\alpha}\frac{1}{U}\,\alpha'
+ \frac{1}{U^{2}}\!\big(\alpha + \beta_{\alpha}\alpha^{3}\big) \nonumber\\
= \frac{2}{\pi\,\mu\,r_{\alpha}^{2}}\,C_M(\tau).
\end{align}

\begin{figure}[!htbp]
    \centering
    \includegraphics[width=0.7\linewidth]{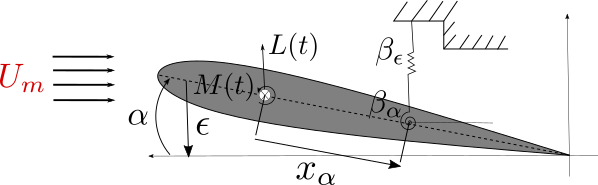}
    \caption{2-DOF aerofoil model with pitch and plunge motion.}
    \label{fig:aerofoil}
\end{figure}

The cubic stiffness terms \(\beta_{\epsilon},\beta_{\alpha}\) capture geometric nonlinearities responsible for LCOs. Figure~\ref{fig:aerofoil} shows a schematic of the model, and Table~\ref{tab:properties} shows the values used for all of the properties used in the simulations.
\begin{table}[!htbp]
    \centering
    \renewcommand{\arraystretch}{1.15}
    \begin{tabular}{lcc}
        \textbf{Parameter} & \textbf{Symbol} & \textbf{Value} \\
        Elastic-axis offset (from mid-chord) & $x_{\alpha}$ & 0.29 \\
        Radius of gyration (about elastic axis) & $r_{\alpha}$ & 0.707 \\
        Mass ratio & $\mu$ & 660.0 \\
        Aerodynamic center offset & $a_h$ & $-0.5$ \\
        Plunge cubic stiffness coefficient & $\beta_{\epsilon}$ & 3.0 \\
        Pitch cubic stiffness coefficient & $\beta_{\alpha}$ & 0 \\
        Pitch damping ratio & $\zeta_{\alpha}$ & 0.03 \\
        Plunge damping ratio & $\zeta_{\epsilon}$ & 0.05 \\
        Frequency ratio (plunge/pitch) & $\omega$ & 0.999 \\
    \end{tabular}
    \caption{Properties of the 2-DOF aerofoil model used for all simulations---expressed in nondimensional form.}
    \label{tab:properties}
\end{table}

The unsteady lift \(C_L\) and moment \(C_M\) coefficients are computed using Theodorsen-type approximations with a {Wagner function}, which accounts for wake-induced memory effects according to
\begin{equation}
\phi(\tau) = 1 - 0.165 e^{-0.0455\tau} - 0.335 e^{-0.3\tau},
\end{equation}
so that
\begin{align}
C_L(\tau) &= 2\pi \!\left[ \alpha(0) + \epsilon'(0) + (0.5-a_h)\alpha'(0) \right]\!\phi(\tau) \nonumber\\
&+ 2\pi \!\int_0^{\tau}\!\!\phi(\tau-\sigma)
\left[\alpha'(\sigma)+\epsilon''(\sigma)+(0.5-a_h)\alpha''(\sigma)\right] d\sigma \nonumber\\
&+ \pi \!\left(\epsilon''-a_h\alpha''+\alpha'\right), \\
C_M(\tau) &= \pi (0.5+a_h)\!\left[ \alpha(0)+\epsilon'(0)+(0.5-a_h)\alpha'(0) \right]\!\phi(\tau) \nonumber\\ 
&+ \pi(0.5+a_h) \nonumber\\
&\!\int_0^{\tau}\!\!\phi(\tau-\sigma)
\left[\alpha'(\sigma)+\epsilon''(\sigma)+(0.5-a_h)\alpha''(\sigma)\right] d\sigma.
\end{align}
The convolution structure implies that current aerodynamic forces depend on the past history of aerofoil motion, introducing intrinsic phase delay and enabling quasiperiodic responses.

\subsection{Stochastic Flow-Speed Excitation Models}
\begin{figure*}[!htbp]
\centering
\includegraphics[width=1\textwidth]{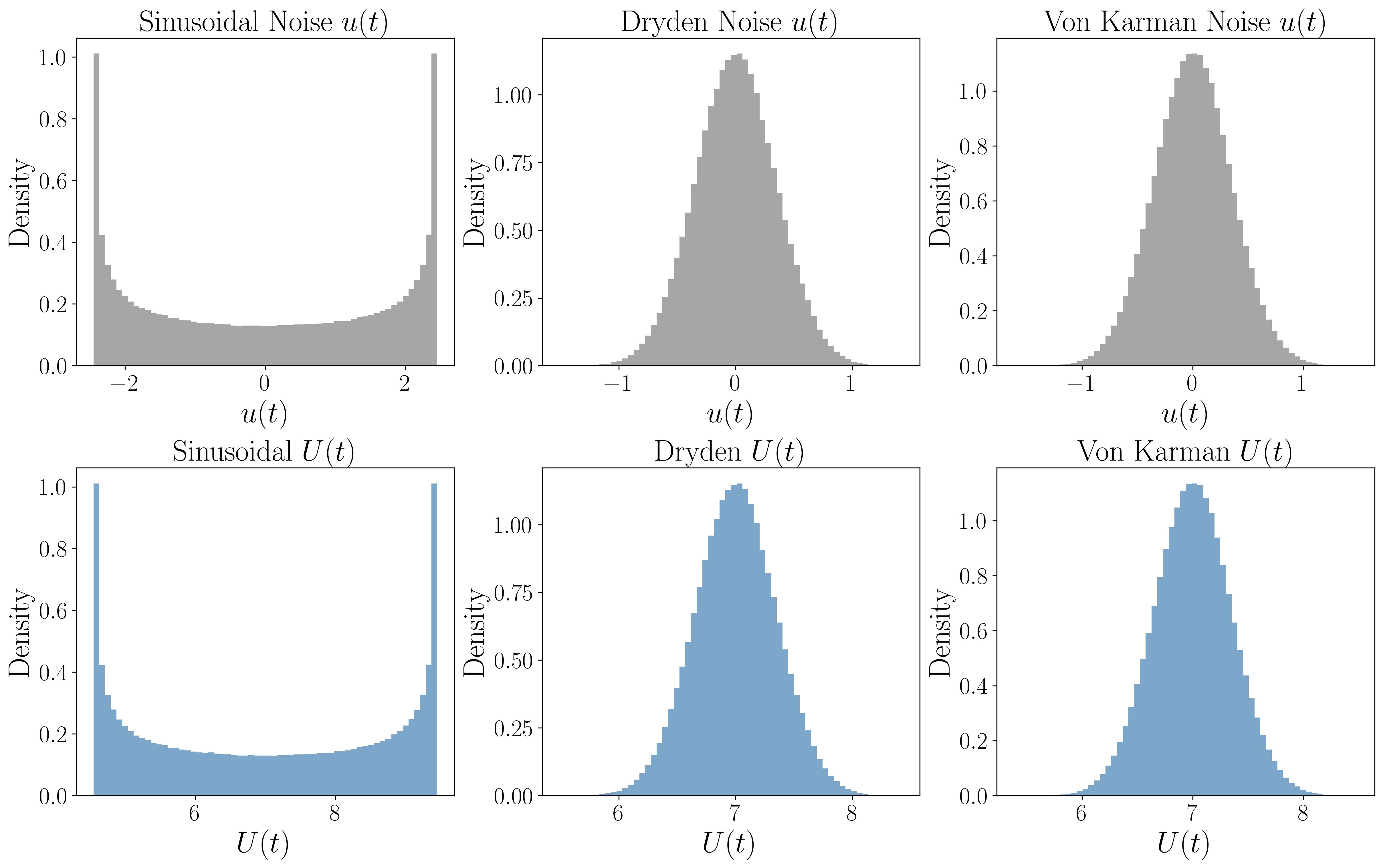}
\caption{Stochastic gust excitation models considered in this study: sinusoidal, Dryden, and von Kármán. The top row shows the probability distributions of the stochastic velocity fluctuations $u(t)$ generated by each excitation model, while the bottom row shows the corresponding distributions of the resulting flow speed $U(t)$ used in the aeroelastic simulations.}
\label{fig:noise}
\end{figure*}

To examine the influence of turbulence spectra on flutter dynamics, the incoming flow velocity is modeled as
\begin{equation}
U(\tau) = U_m + \sigma \Delta U(\tau),
\end{equation}
where \(U_m\) is the mean flow speed and \(\Delta U(\tau)\) represents stochastic fluctuations.  
In addition to the sinusoidal perturbation model considered in the literature \cite{Venkatramani2018}, we study Dryen and von Kármán turbelence models. 
These three excitation models---shown in Fig.~\ref{fig:noise}---are described below, and were selected to represent distinct classes of stochastic input.

\paragraph{(a) Sinusoidal perturbation:}
\begin{equation}
\Delta U = U_m \sin(\omega_r \tau)
\end{equation}
where \( \omega_r \) is a fluctuating frequency modeled as \( \omega_r = \omega_1 + \kappa R(\tau) \), with \( R(\tau) \) being a uniform random variable. 

\paragraph{(b) Dryden turbulence model~\cite{Beal1993}:}
The velocity \(\Delta U = u(t)\) is generated by the first-order shaping filter
\begin{equation}
\dot{u}(t) = -\frac{U_m}{L_u}u(t) + \sqrt{\frac{2\sigma_u^2 U_m}{L_u}}\,\eta(t),
\end{equation}
where \(L_u\) is the turbulence length scale and \(\sigma_u^2\) its variance.  The resulting spectrum decays as \(S(f)\!\propto\!(1+L_u^2 f^2)^{-1}\), producing smoothly correlated fluctuations.

\paragraph{(c) von Kármán turbulence model~\cite{Beal1993}:}
A second-order shaping filter approximates the von Kármán spectrum,
\begin{equation}
\ddot{u}(t) + 2\zeta_u\omega_u \dot{u}(t) + \omega_u^2 u(t)
   = \sqrt{2\sigma_u^2\omega_u^3}\,\eta(t),
\end{equation}
with damping ratio \(\zeta_u\) and break frequency \(\omega_u = U_m/L_u\).  Compared with the Dryden model, it yields a steeper high-frequency roll-off and more physically realistic coherence of turbulence eddies.

\section{Methods}

We assume that the system is ergodic, allowing the stationary distribution to be approximated via ensemble averaging over monte carlo realizations. We use homological bifurcation analysis to detect phenomenological bifurcations in stochastic systems~\cite{Tanweer2024} and apply it to the four-dimensional Kernel Density Estimates (KDEs) estimated from the state-space response using 100 monte carlo simulations per mean speed $U_m \in [2, 10]$. The following sections describe the homological bifurcation analysis framework, while the bandwidth estimation used for the KDEs is described in Appendix~\ref{app:kde_unsupervised}. 

\subsection{Superlevel persistence of the probability density function}

Given a unit-normalized stationary KDE $p(\bm{x})$ of the aerofoil state vector $\bm{x} = [\alpha, \epsilon, \dot{\alpha}, \dot{\epsilon}]$, we construct a \emph{superlevel filtration} of cubical complexes (see~\cite{Tanweer2024} or Appendix~\ref{app:topological-background} for more details on this)
\[
K_{L_1} \subseteq K_{L_2} \subseteq \cdots \subseteq K_{L_N},
\quad \text{where } K_L = p^{-1}([L,\infty)).
\]
Each cubical complex $K_L$ represents the subset of phase space where the KDE exceeds a probability threshold $L$. As $L$ decreases from its maximum value to zero, high-probability regions merge and new topological features (connected components, loops, or voids) appear and disappear. Persistent homology quantifies these topological features by tracking the birth and death of homology classes as a function of $L$.

For the aeroelastic model, we compute persistence over $H_0$, $H_1$, and $H_2$ homology groups corresponding to connected components, limit-cycle loops, and voids, respectively.

\subsection{Homological bifurcation plots}

A family of KDEs $\{p_i(\bm{x})\}$ is generated by varying the bifurcation parameter---here, the mean flow speed $U_m$. For each $p_i(\bm{x})$, the persistence computation yields a vector of Betti numbers
\[
{\boldsymbol{\beta}}_i = [\beta_i(L_1), \beta_i(L_2), \ldots, \beta_i(L_N)],
\]
where $i$ values of $0$, $1$, and $2$ give the vectors  $\boldsymbol{\beta}_0$, $\boldsymbol{\beta}_1$, and $\boldsymbol{\beta}_2$, respectively. 
These Betti vectors are then stacked over the range of $U_m$ values to form a two-dimensional map showing how each homology group evolves across both the filtration parameter and the bifurcation parameter.

\begin{figure*}[!htbp]
\centering
\includegraphics[width=0.8\textwidth]{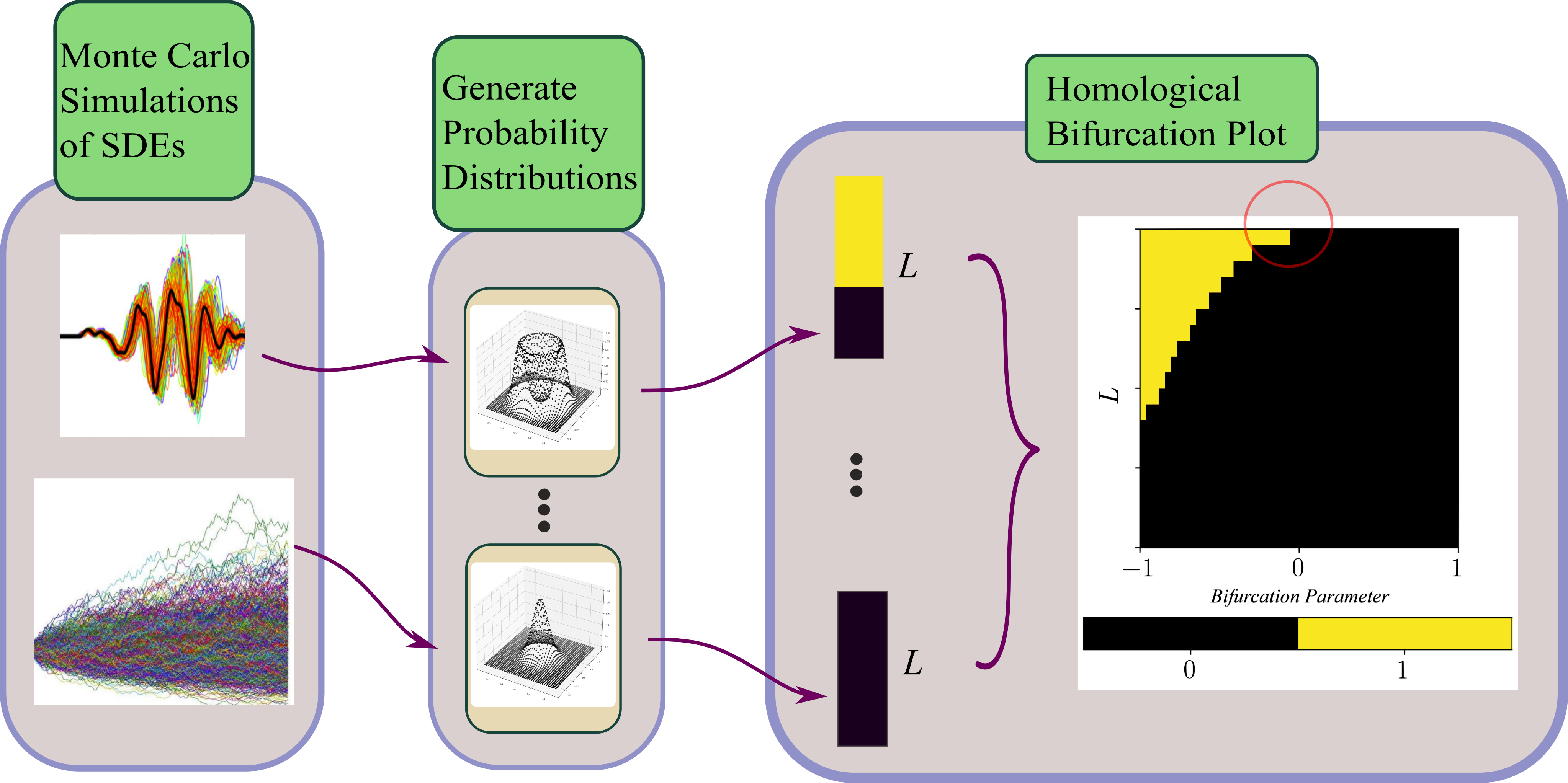}
\caption{Schematic overview of the homological bifurcation analysis framework. 
Monte Carlo simulations generate ensembles of trajectories which are used to estimate the distribution. Superlevel filtration of cubical complexes is computed for each KDE, and the resulting Betti vectors yield the homological bifurcation plot, }
\label{fig:method}
\end{figure*}

The resulting heatmap---referred to as a \textit{homological bifurcation plot}~\cite{Tanweer2024}---provides a quantitative visualization of changes in the topology of the KDE. Abrupt transitions in the rank of a homology group $\beta_p$ indicate a qualitative shift in the system’s probabilistic structure and thus mark a P-bifurcation. See Fig.~\ref{fig:method} for a schematic of the method.

\section{Results and Discussion}
\label{sec:results}

The stochastic equations were integrated for a range of mean flow speeds \(U_m\) for each turbulence model and the long-term system response 
\[
\mathbf{x}(t) = (\epsilon, \alpha, \dot{\epsilon}, \dot{\alpha})
\]
was sampled to reconstruct the four-dimensional state-space probability density function using kernel density estimation. The density was normalized to unit total probability, and its topological evolution was examined using homological bifurcation plots over the normalized filtration parameter \( \rho \in [0,1] \). 

We first examine the direct dynamical response of the aeroelastic system in the time domain and in low-dimensional phase-space  projections in Section~\ref{sec:timeDomain} followed by the topological analysis in Section~\ref{sec:hombifPlots}. 

\subsection{Time-Domain and Phase-Space Response Across Flow Speeds}
\label{sec:timeDomain}

\begin{figure*}[!htbp]
\centering
\includegraphics[width=0.3\linewidth]{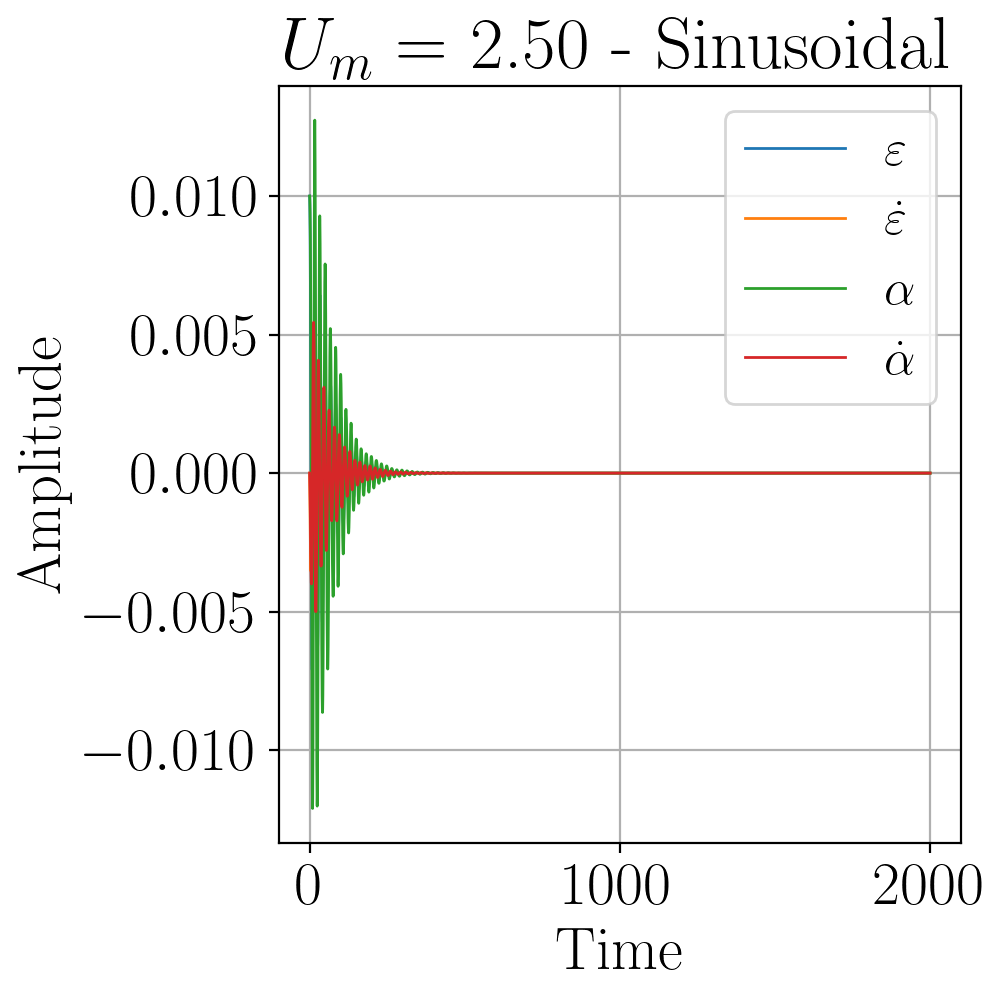}
\includegraphics[width=0.3\linewidth]{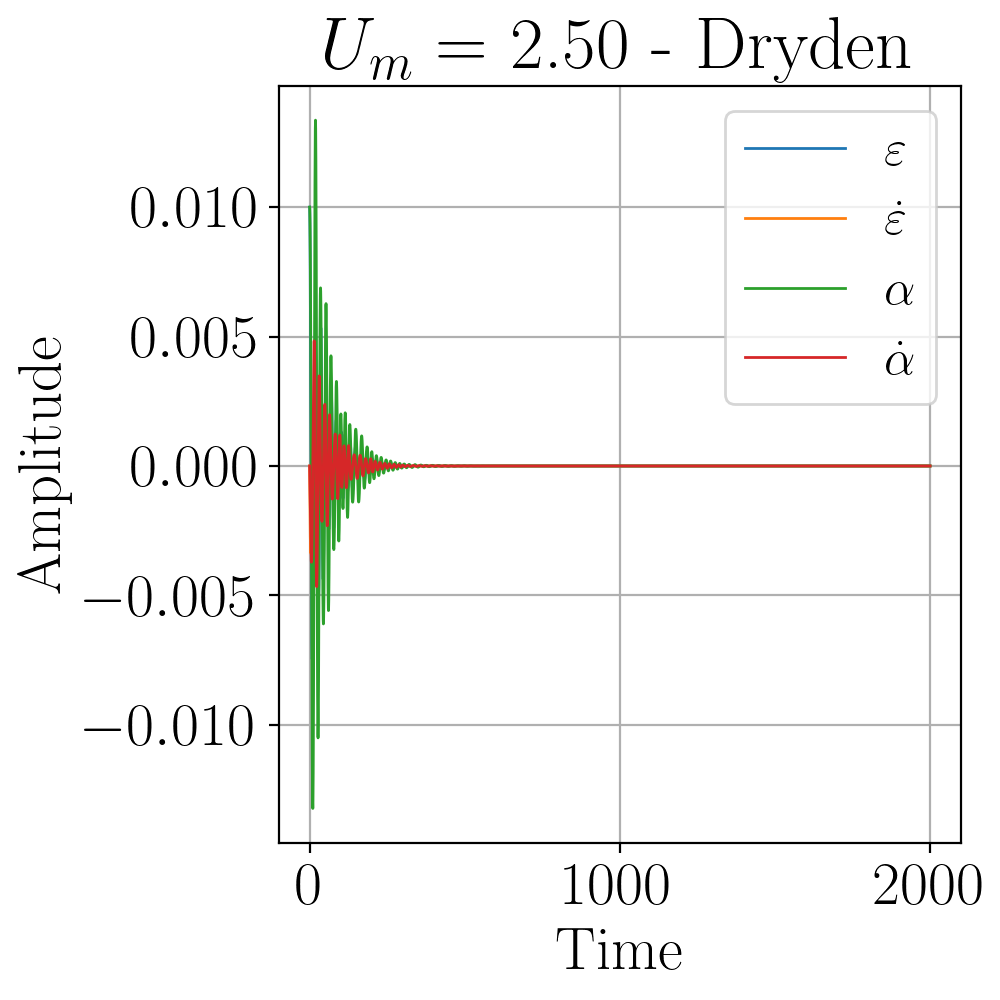}
\includegraphics[width=0.3\linewidth]{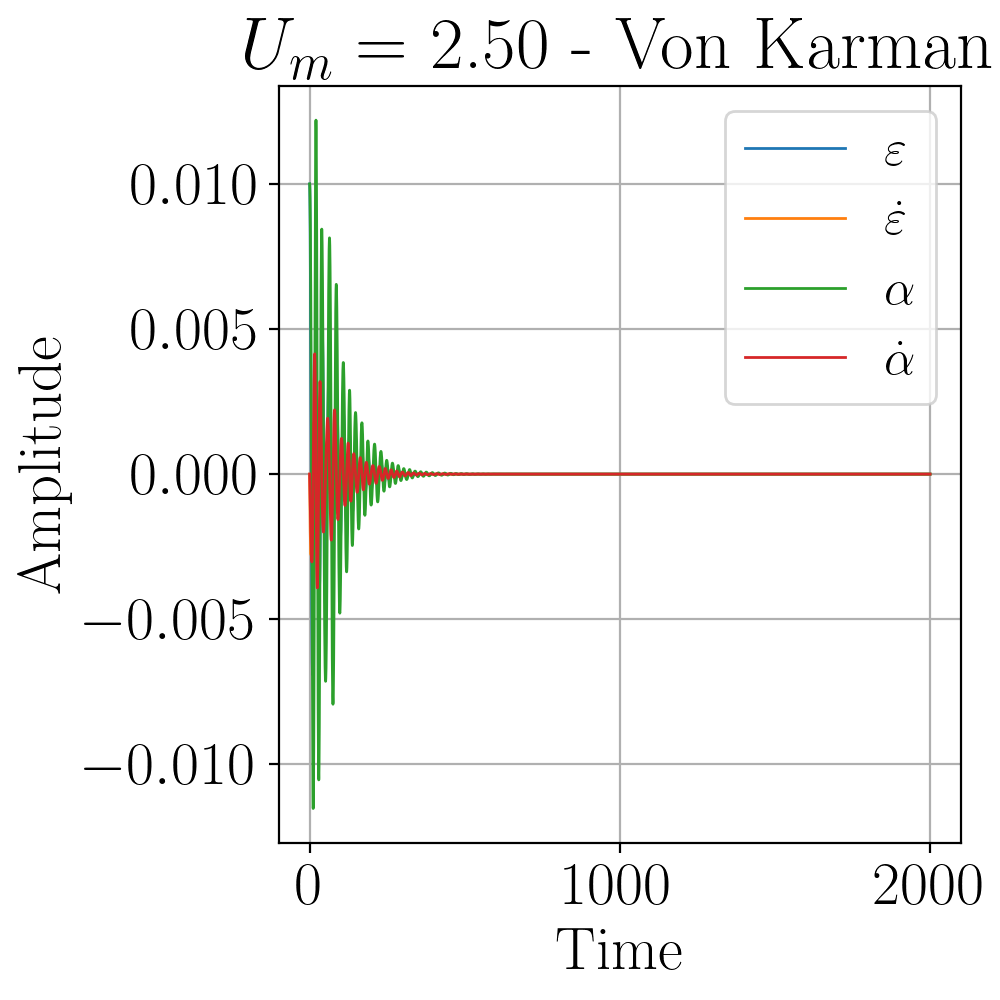}

\includegraphics[width=0.3\linewidth]{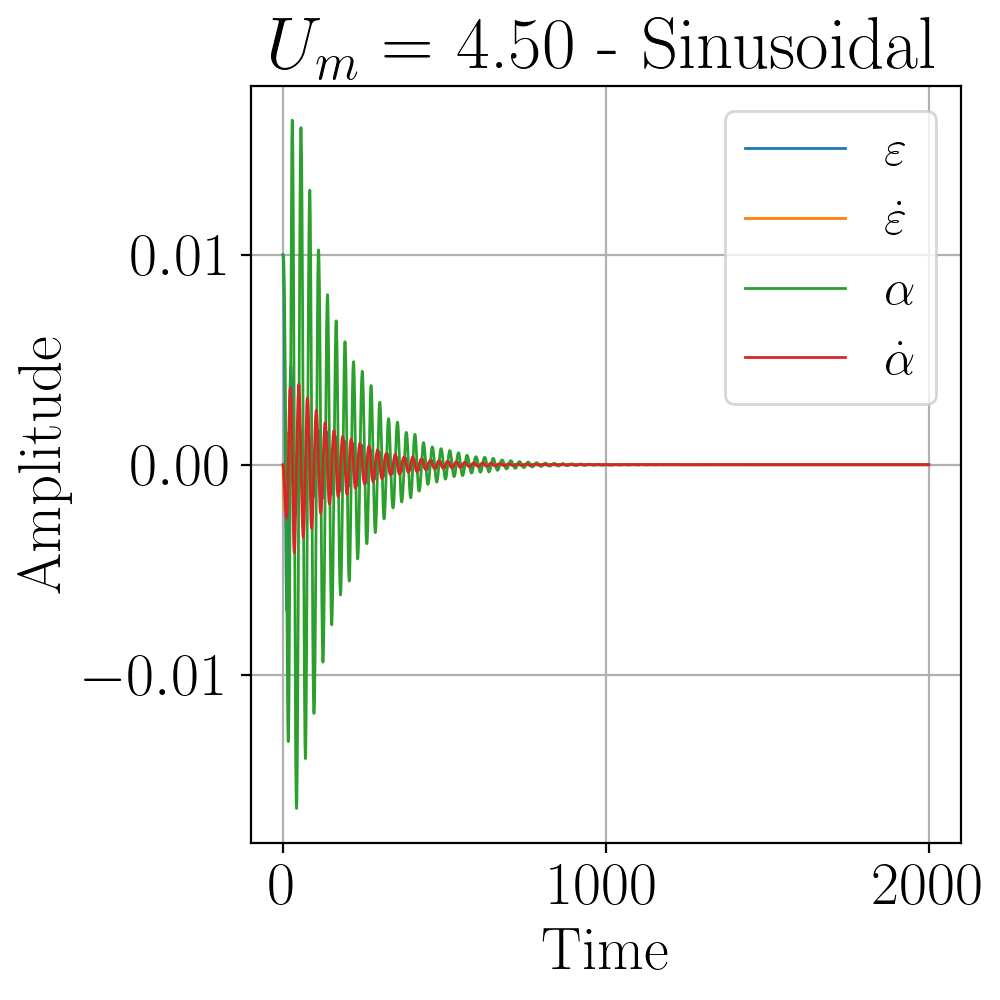}
\includegraphics[width=0.3\linewidth]{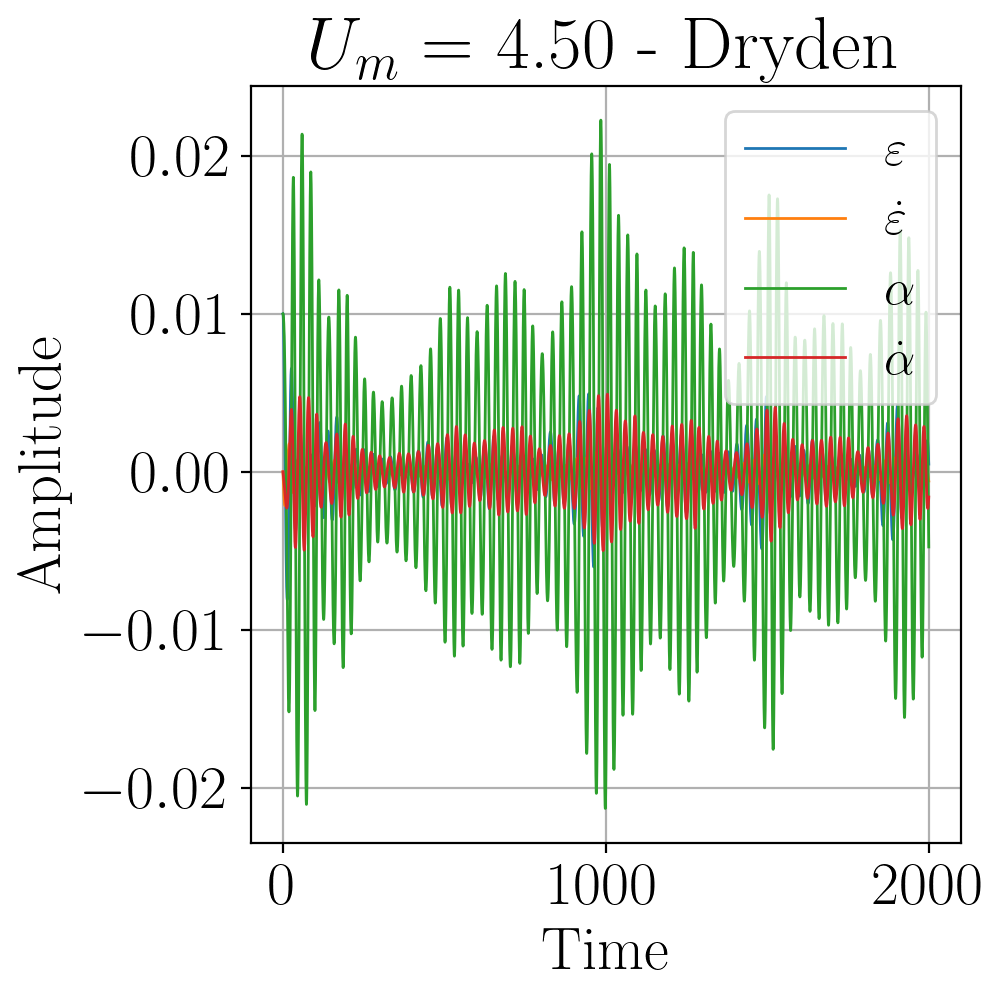}
\includegraphics[width=0.3\linewidth]{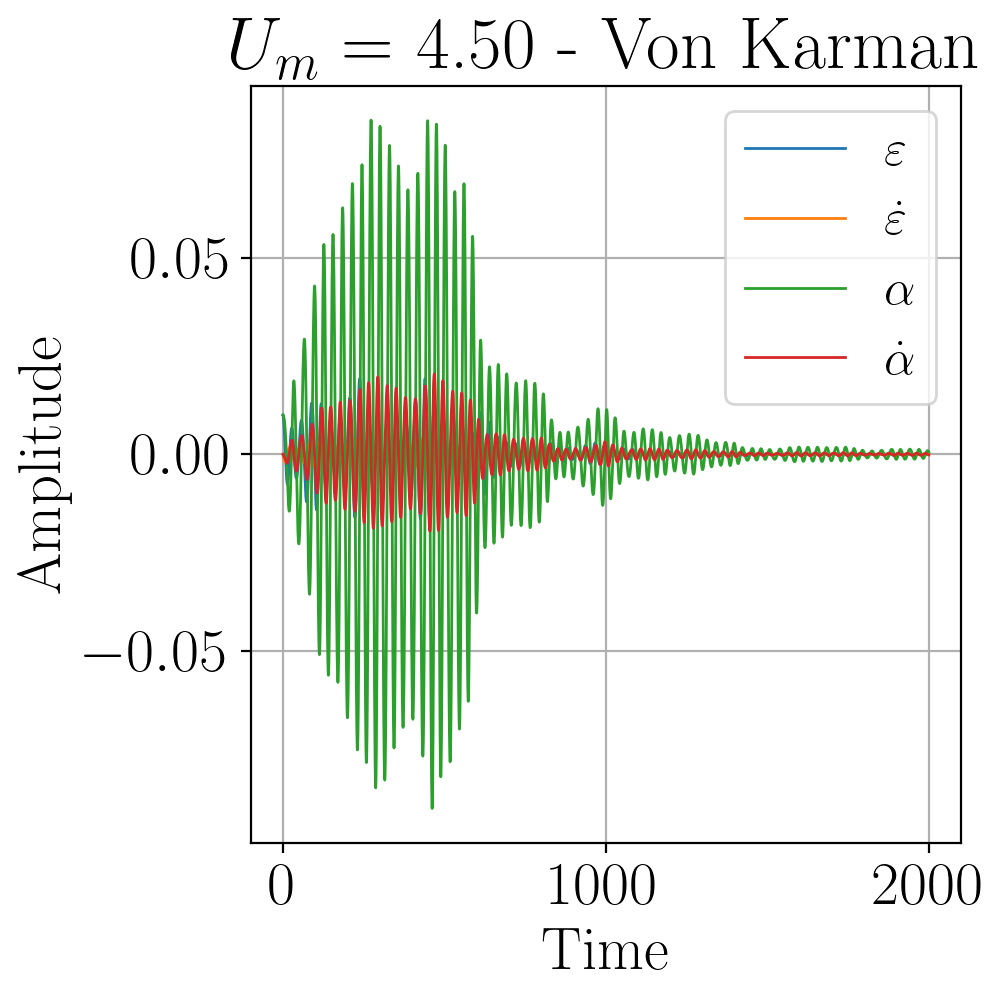}

\includegraphics[width=0.3\linewidth]{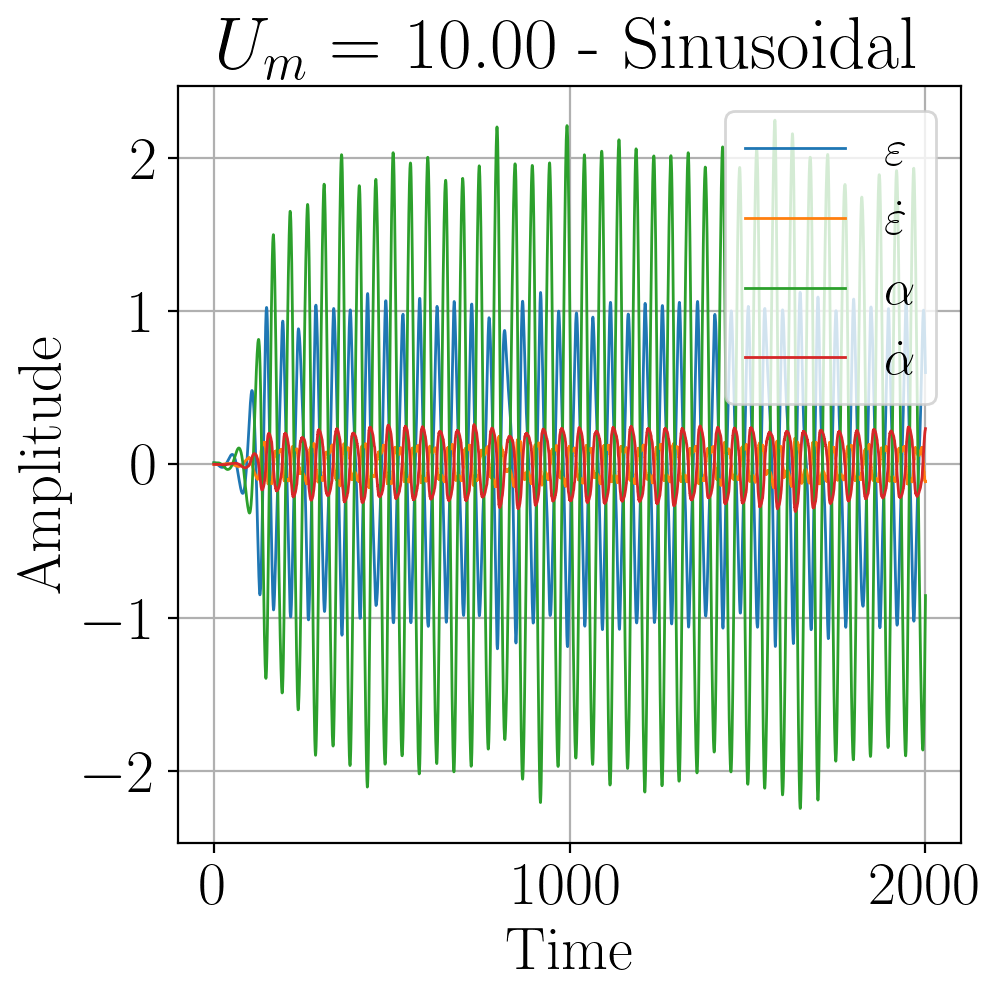}
\includegraphics[width=0.3\linewidth]{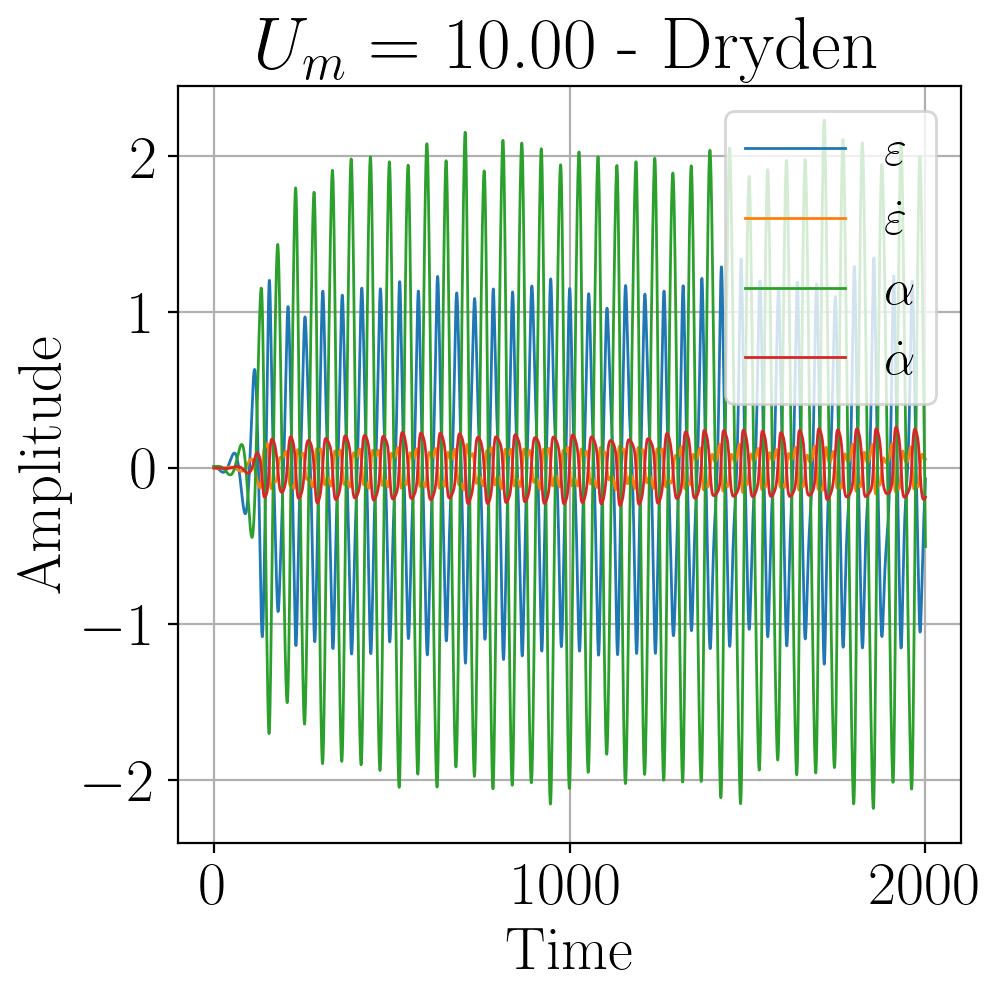}
\includegraphics[width=0.3\linewidth]{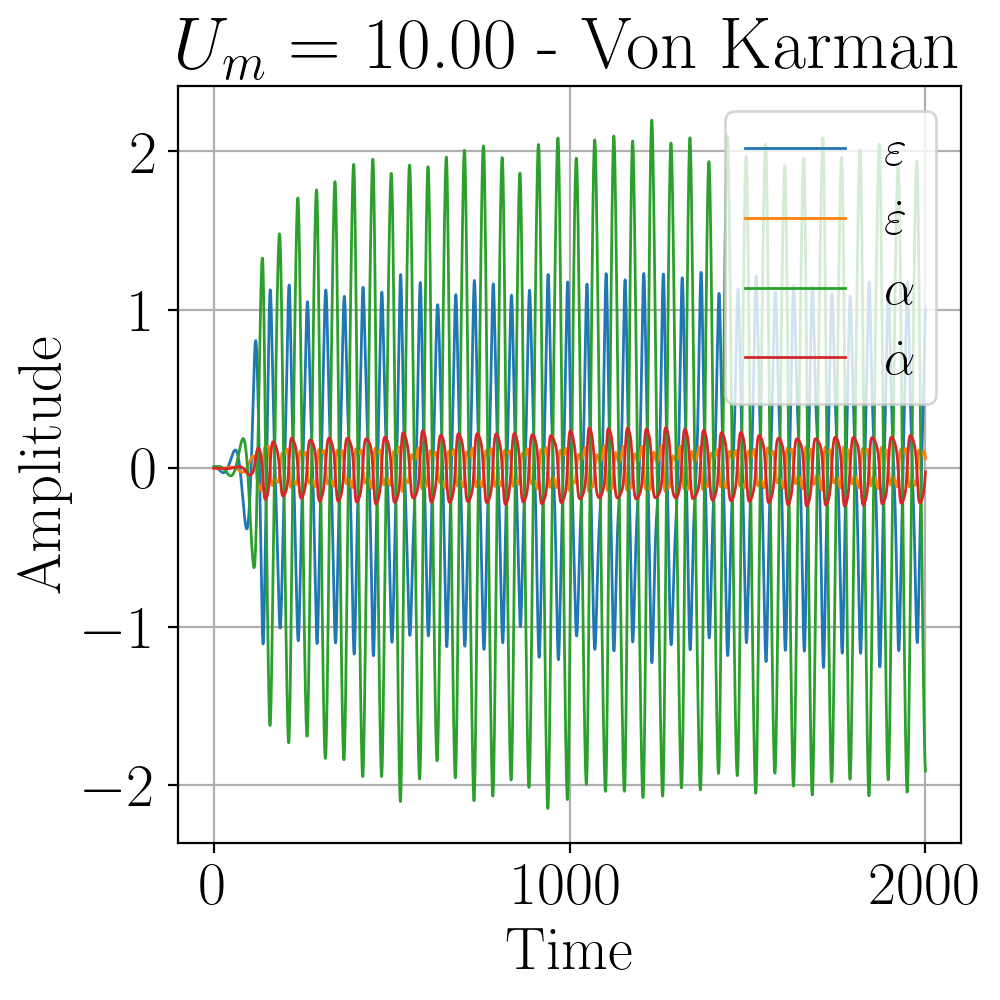}
\caption{Sample time-series (one of the many montecarlo simulations) of pitch $\alpha(t)$ and plunge $\epsilon(t)$ with their velocities, at subcritical, near-critical, and supercritical flow speeds for the sinusoidal, Dryden, and von Kármán excitation models.}
\label{fig:timeseries_all}
\end{figure*}

\begin{figure*}[!htbp]
    \centering
    \includegraphics[width=0.3\linewidth]{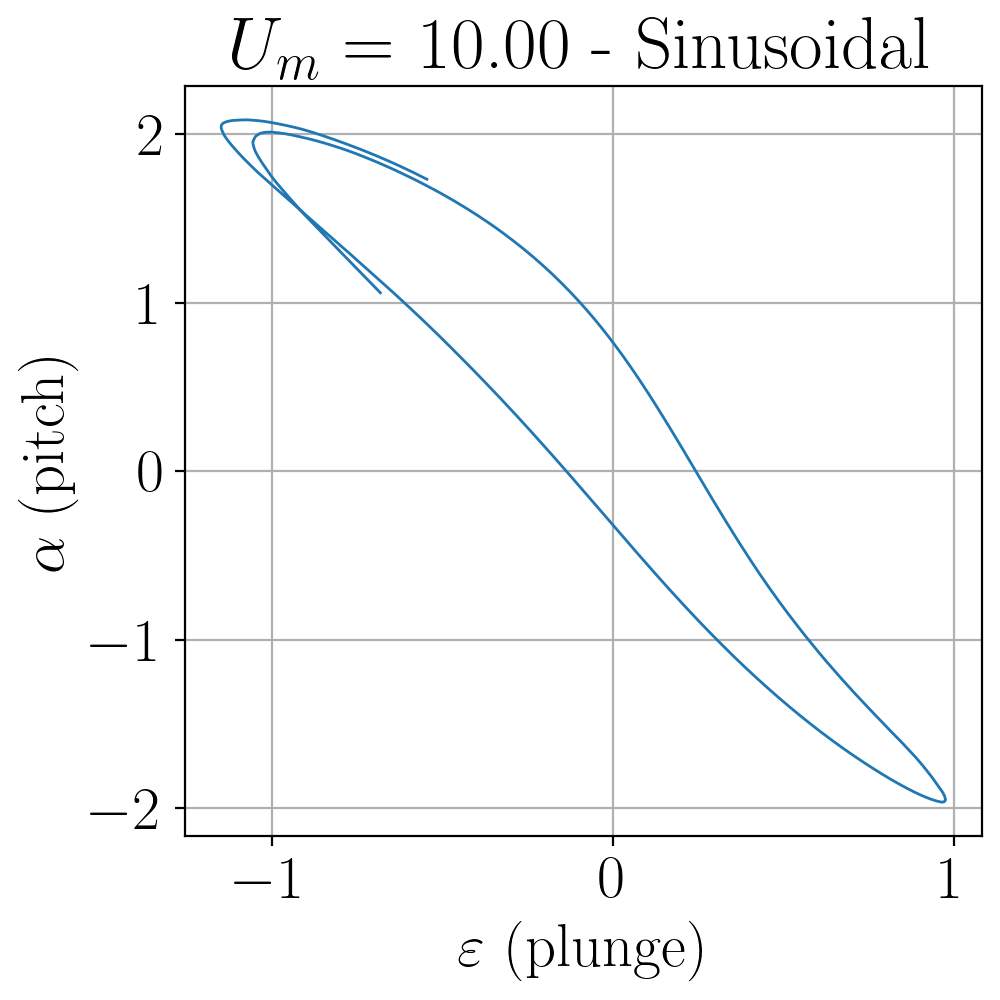}
    \includegraphics[width=0.3\linewidth]{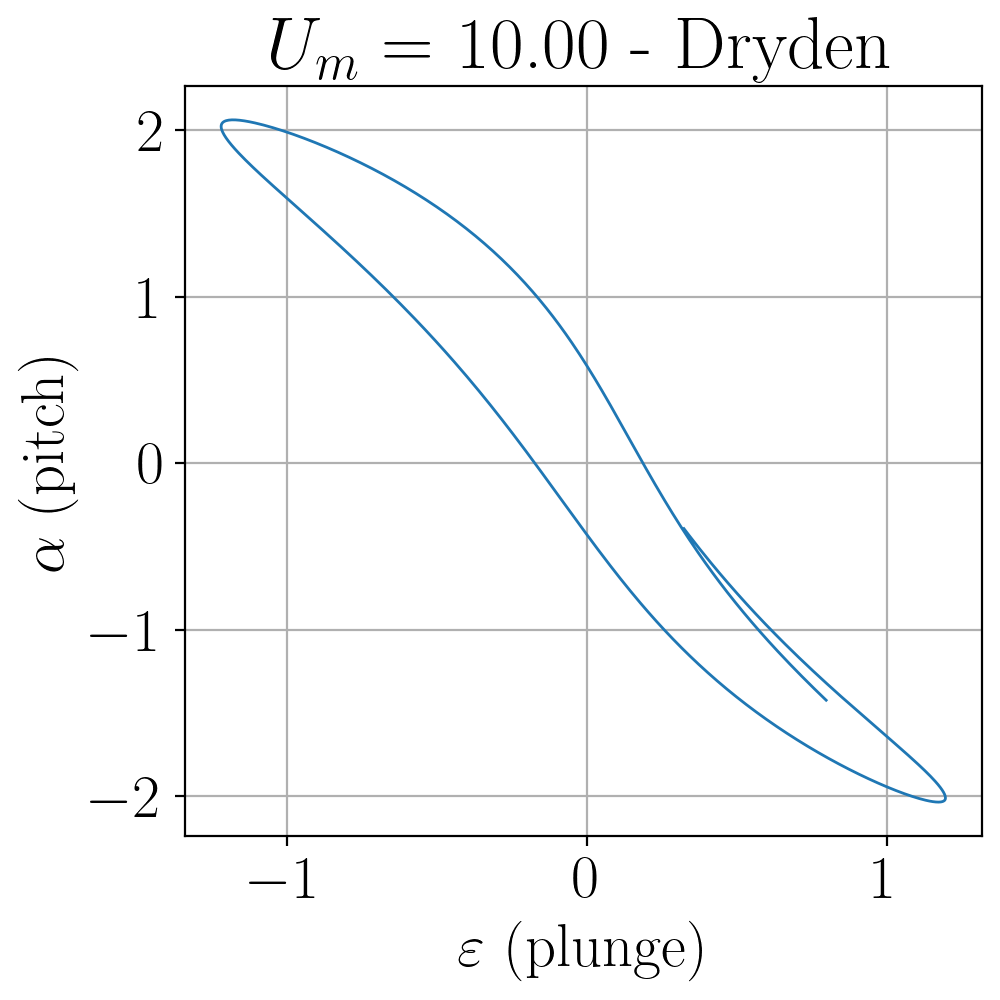}
    \includegraphics[width=0.3\linewidth]{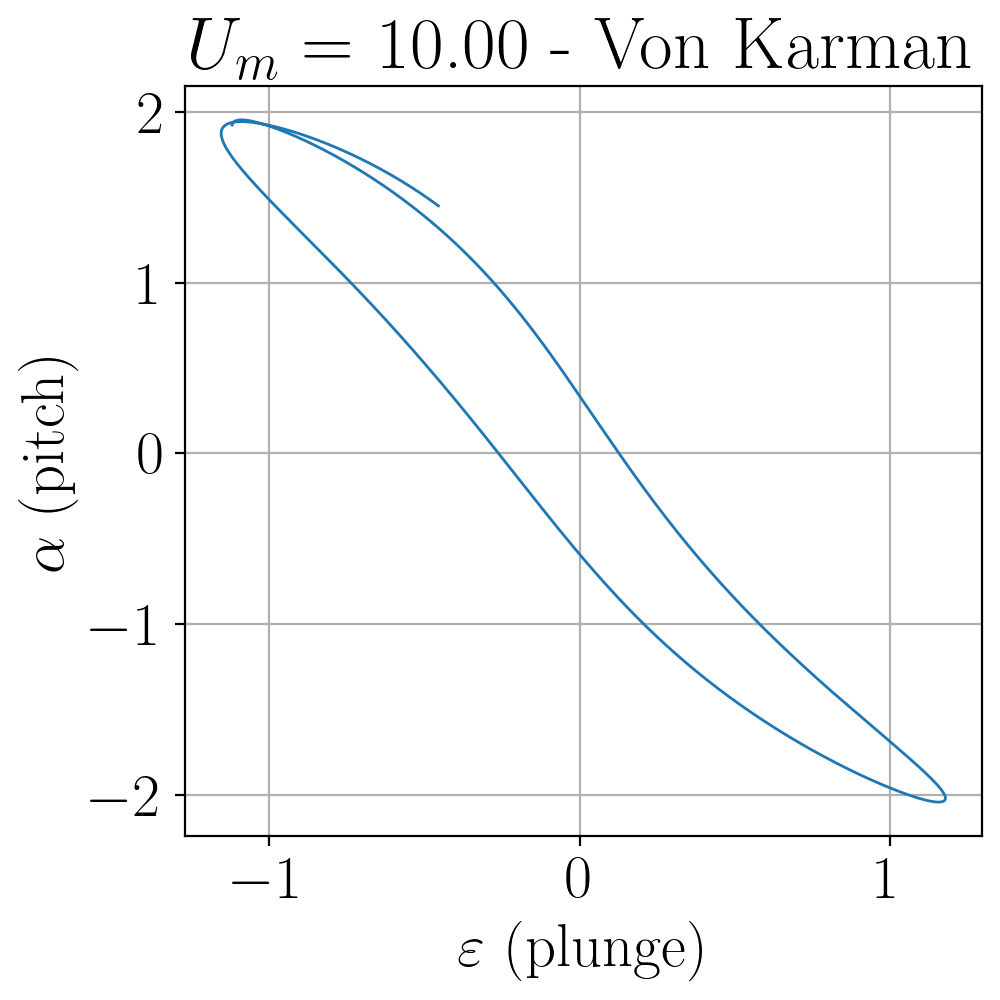}
    
    \includegraphics[width=0.3\linewidth]{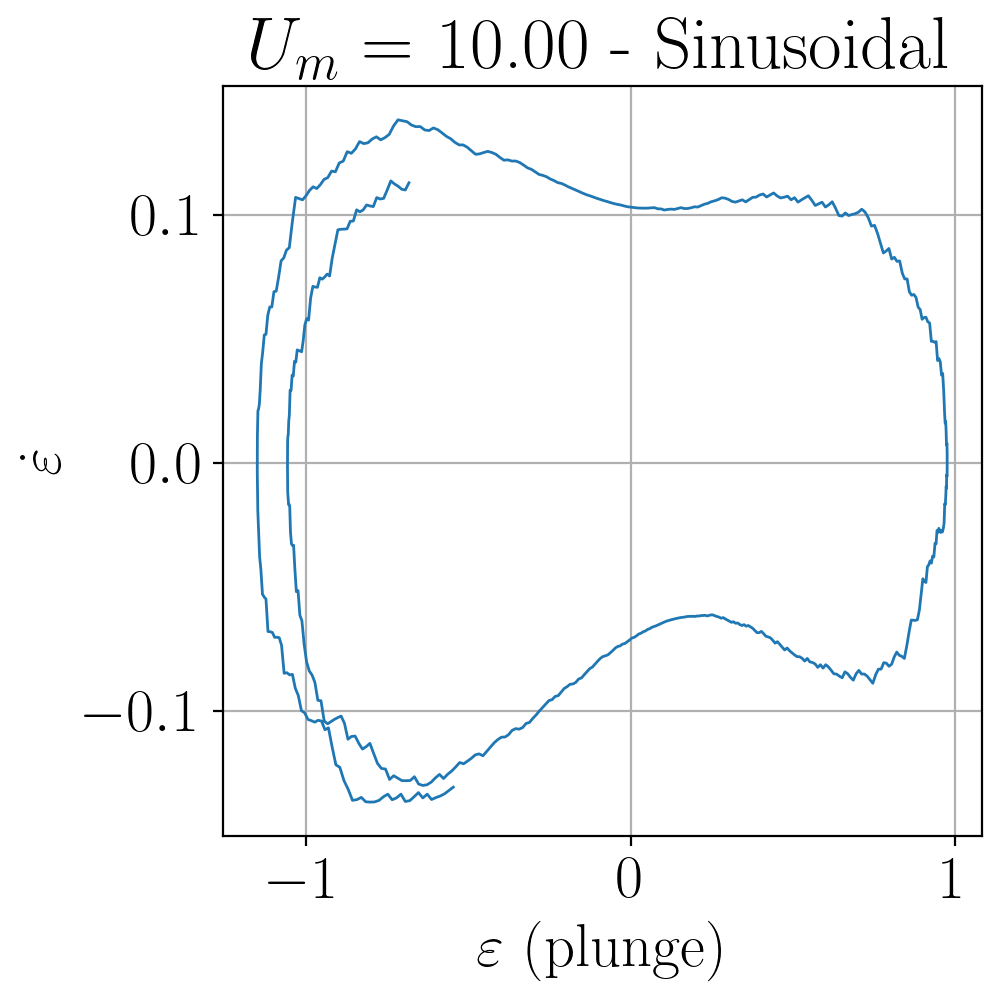}
    \includegraphics[width=0.3\linewidth]{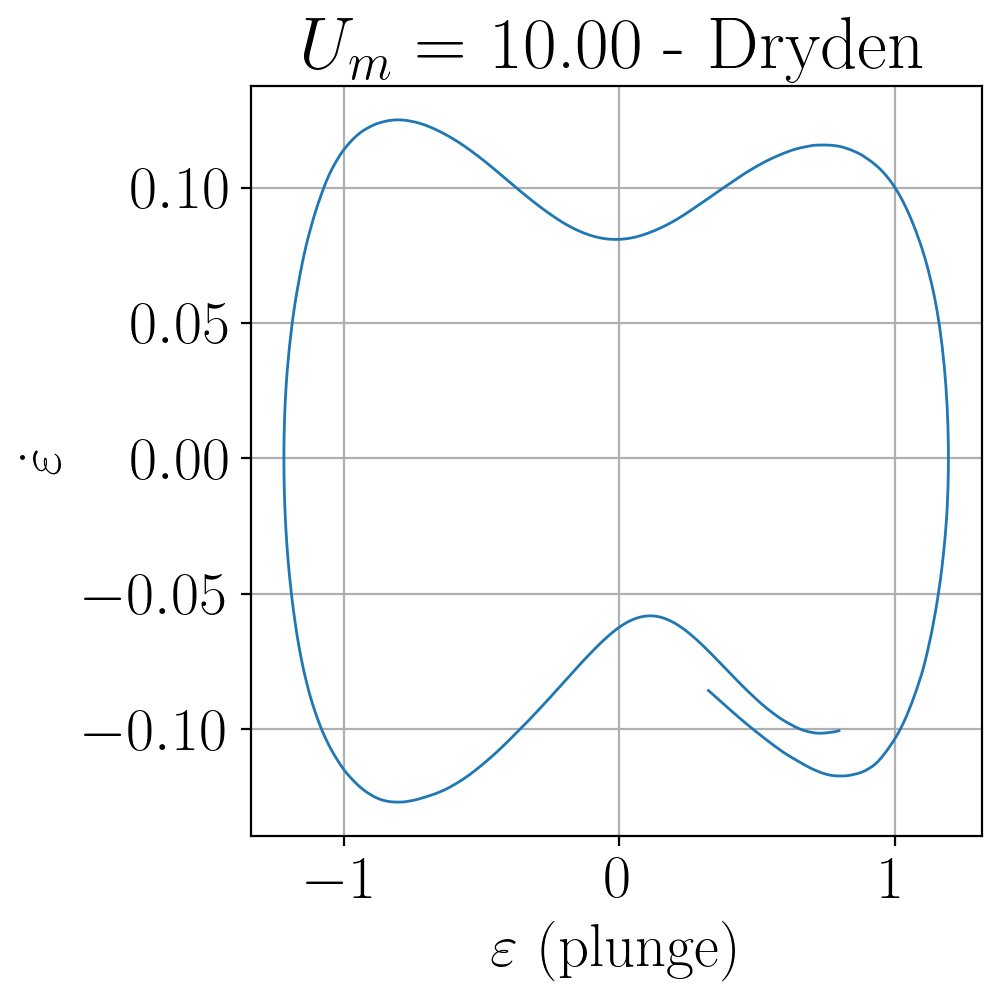}
    \includegraphics[width=0.3\linewidth]{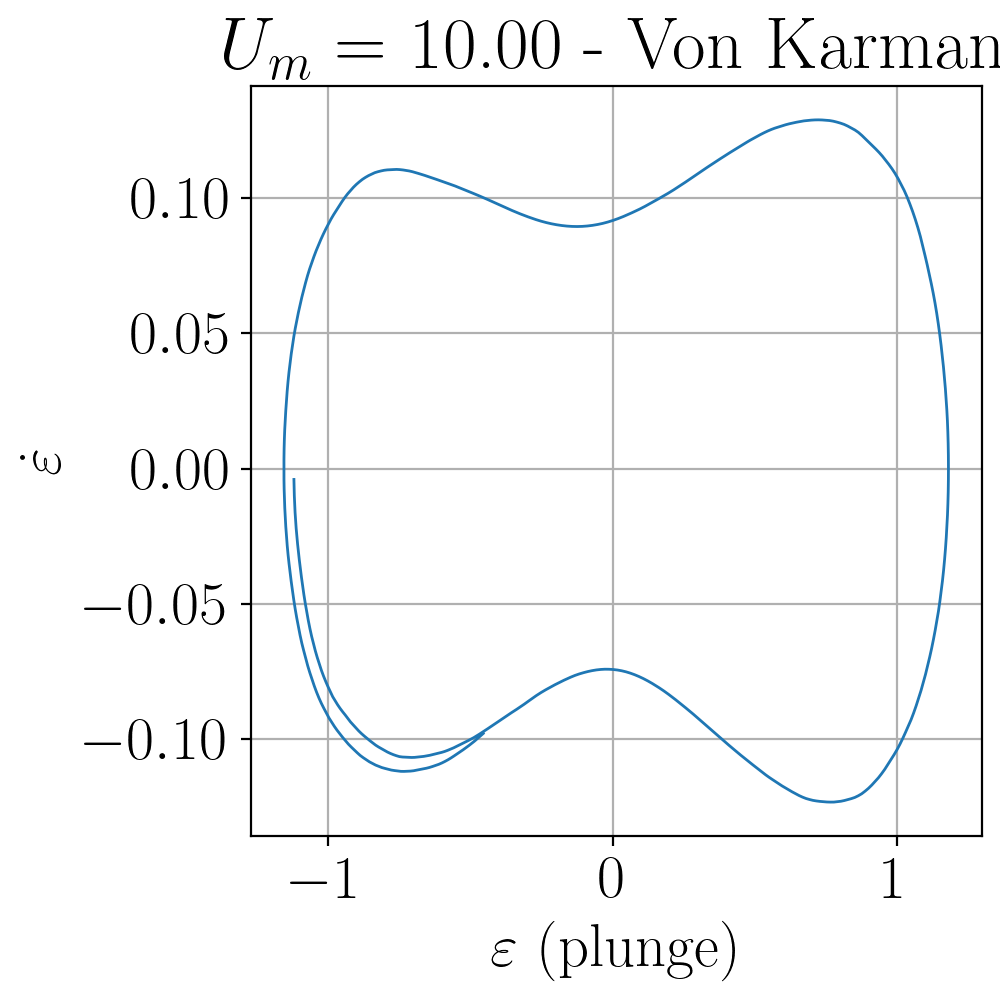}

    \includegraphics[width=0.3\linewidth]{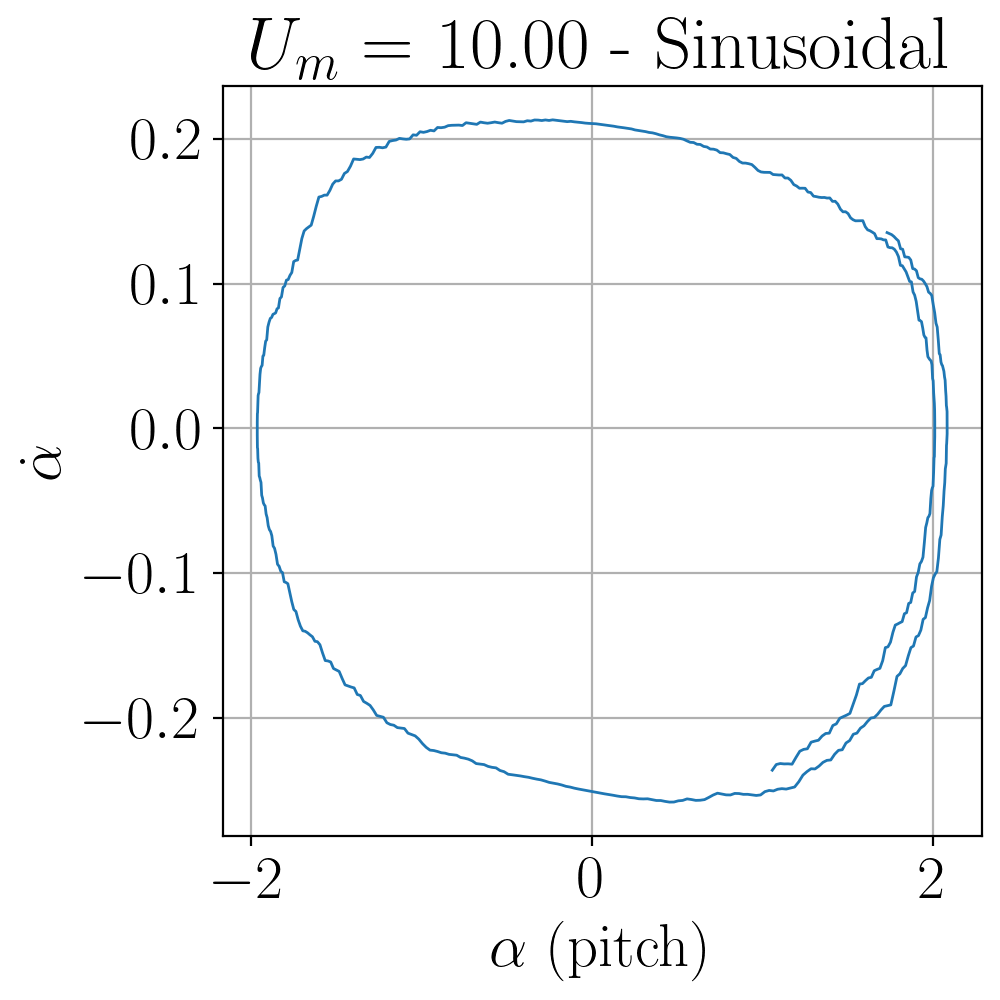}
    \includegraphics[width=0.3\linewidth]{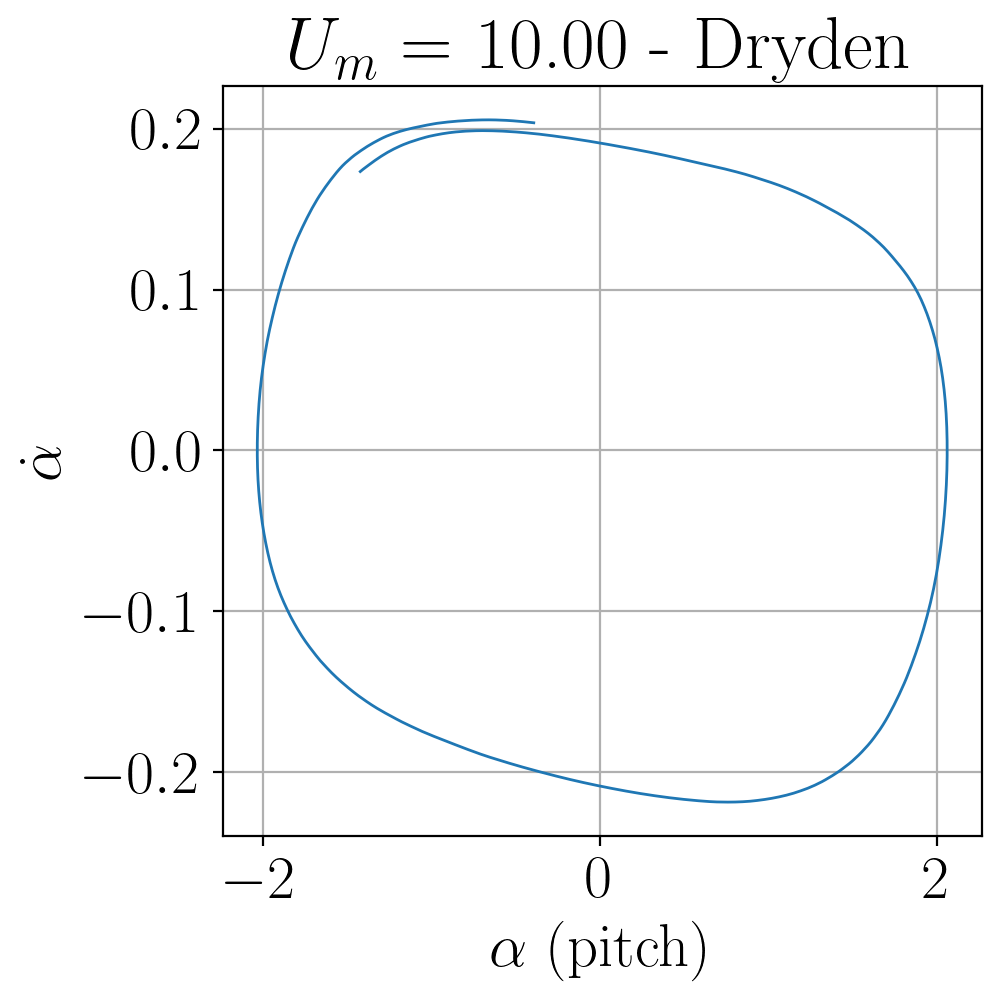}
    \includegraphics[width=0.3\linewidth]{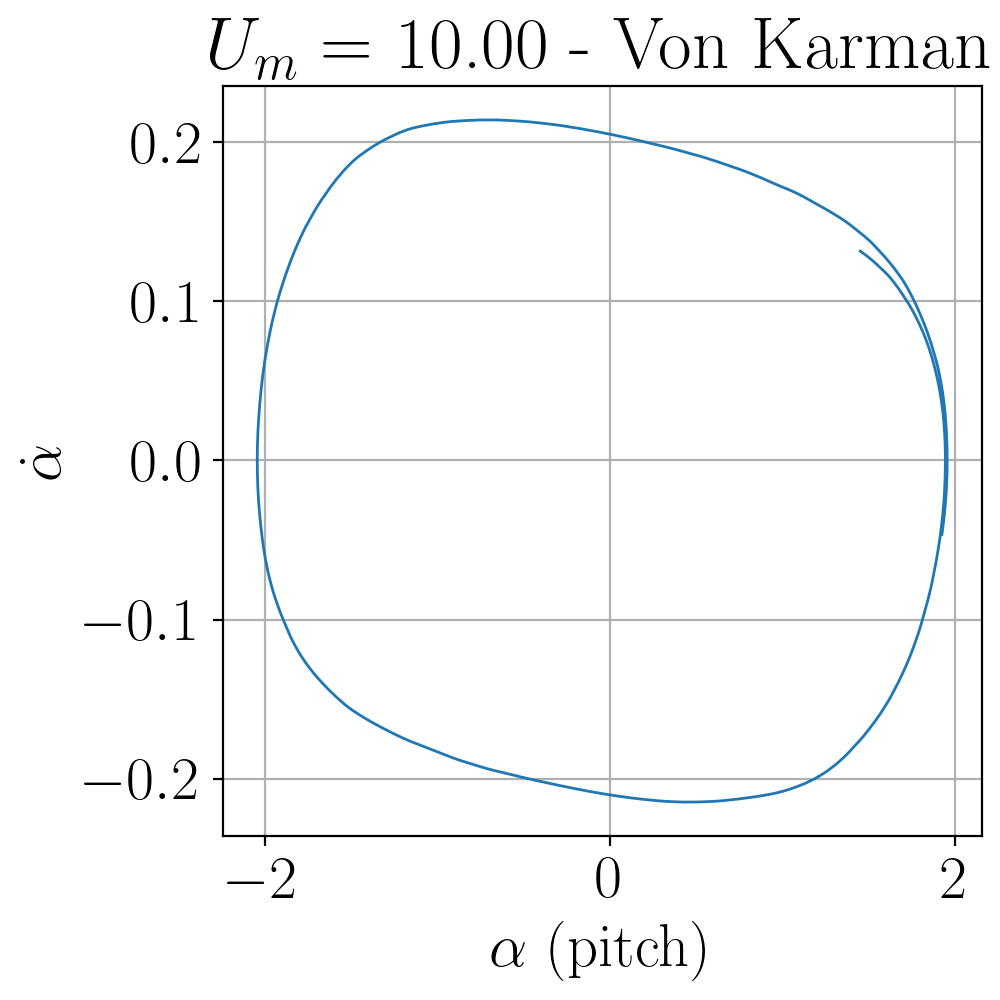}
    \caption{Sample phase-space projections (one of the many montecarlo simulations) $(\alpha,\varepsilon)$, $(\varepsilon,\dot{\varepsilon})$ and $(\alpha,\dot{\alpha})$ showing limit cycles for the three noise models at a speed of 10 m/s. The limit cycles are very narrow in $(\alpha,\varepsilon)$ space, but wider in the other two.}
    \label{fig:phase_all}
\end{figure*}

Figures~\ref{fig:timeseries_all} and \ref{fig:phase_all} summarize representative 
time histories and phase portraits for three mean-flow speeds---a subcritical regime 
($U_m = 2.5$), a near-critical regime ($U_m = 4.5$), and a clearly supercritical regime 
($U_m = 10.0$).

At low flow speeds ($U_m = 2.5$), all three models exhibit strongly damped transient responses. 
The pitch and plunge displacements, along with their velocities, decay to zero. Dryden and von Kármán 
excitations introduce small stochastic perturbations, but these remain well within the linear decay envelope too. The corresponding phase-space trajectories collapse to a single point, reflecting a unimodal stationary density. As the flow speed approaches the instability threshold ($U_m = 4.5$), differences among the three 
models begin to emerge. The sinusoidal excitation still produces a decaying response, with no growth 
in amplitude. In contrast, both Dryden and von Kármán turbulence inject sufficient correlated energy to generate intermittent bursts of oscillation. These appear as modulated time-series 
envelopes and as broadened, partially closed loops in the $(\varepsilon,\alpha)$, $(\varepsilon,\dot{\varepsilon})$, and $(\alpha,\dot{\alpha})$ projections---although the system remains nominally stable due to the low amplitude of these. For supercritical flow speeds ($U_m = 10.0$), all three excitation models lead to sustained large-amplitude oscillations in pitch and plunge. 

Overall, the time-domain and phase-space responses confirm that turbulence models with temporal correlation (Dryden and von Kármán) accelerate the onset of flutter-like oscillatory behavior and yield richer oscillatory structure above the critical flow speed, a trend that we quantitatively capture using the topological signatures presented in the following sections.

\subsection{Homological Bifurcation Plots}
\label{sec:hombifPlots}

\begin{figure*}[!htbp]
\centering
\includegraphics[width=0.32\textwidth]{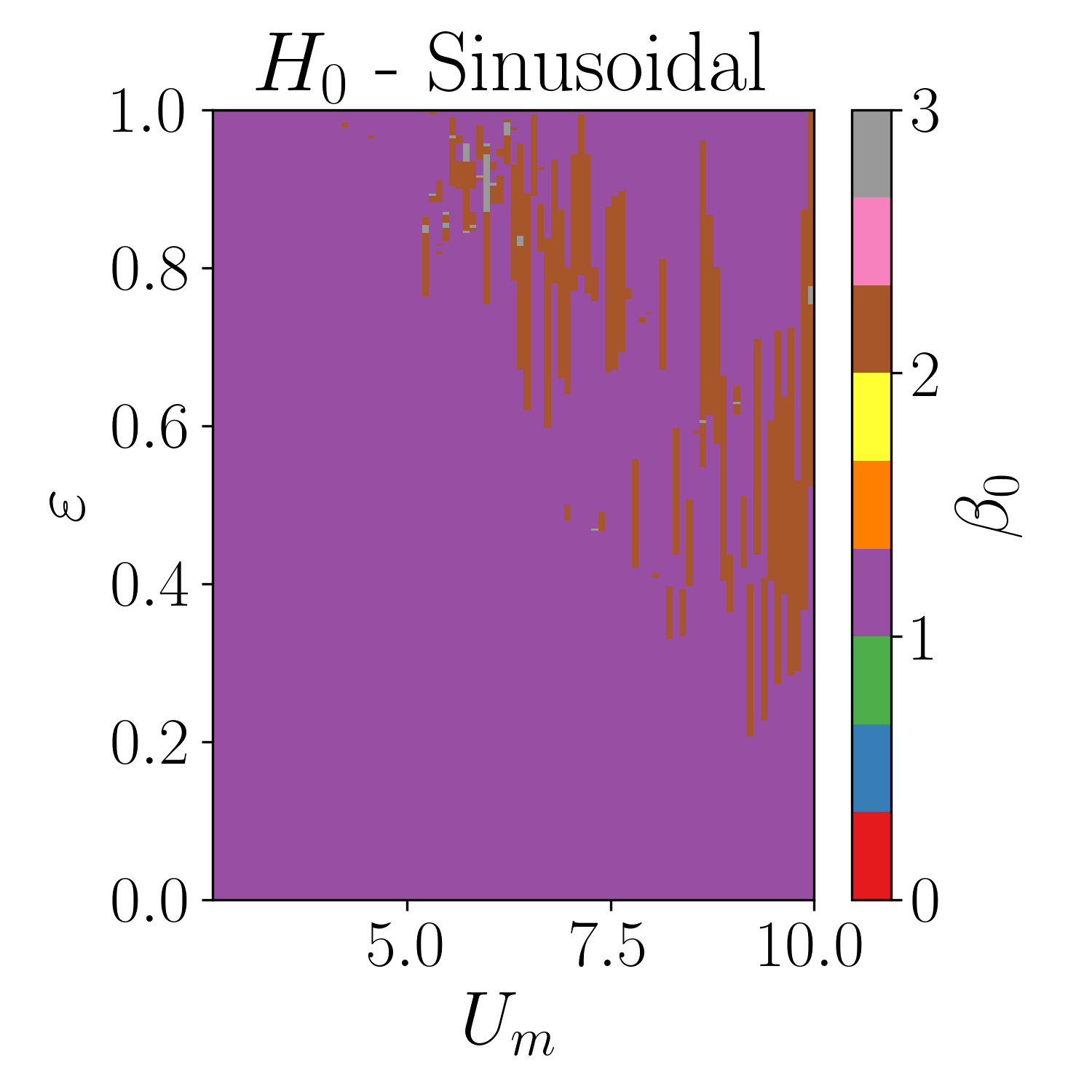}
\includegraphics[width=0.32\textwidth]{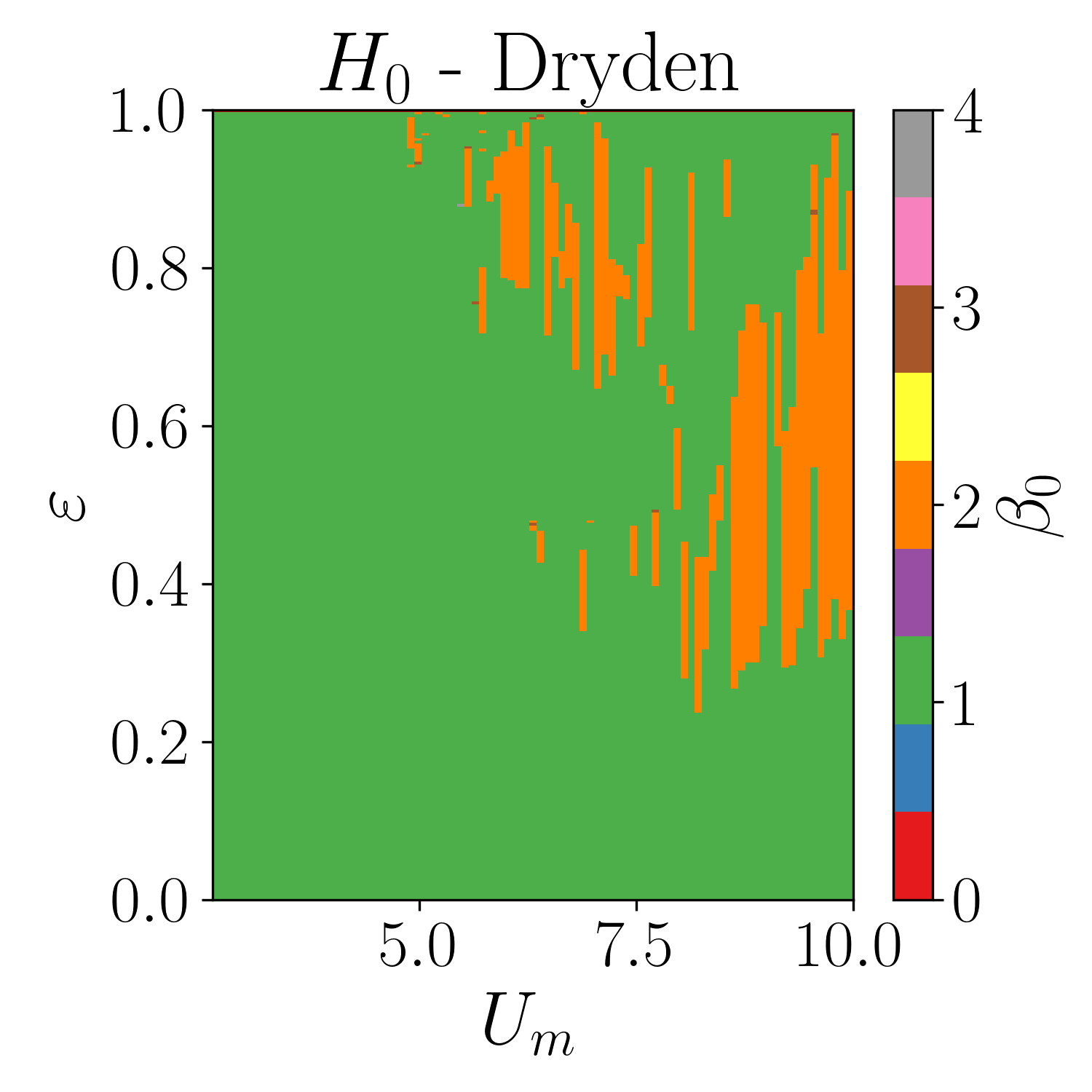}
\includegraphics[width=0.32\textwidth]{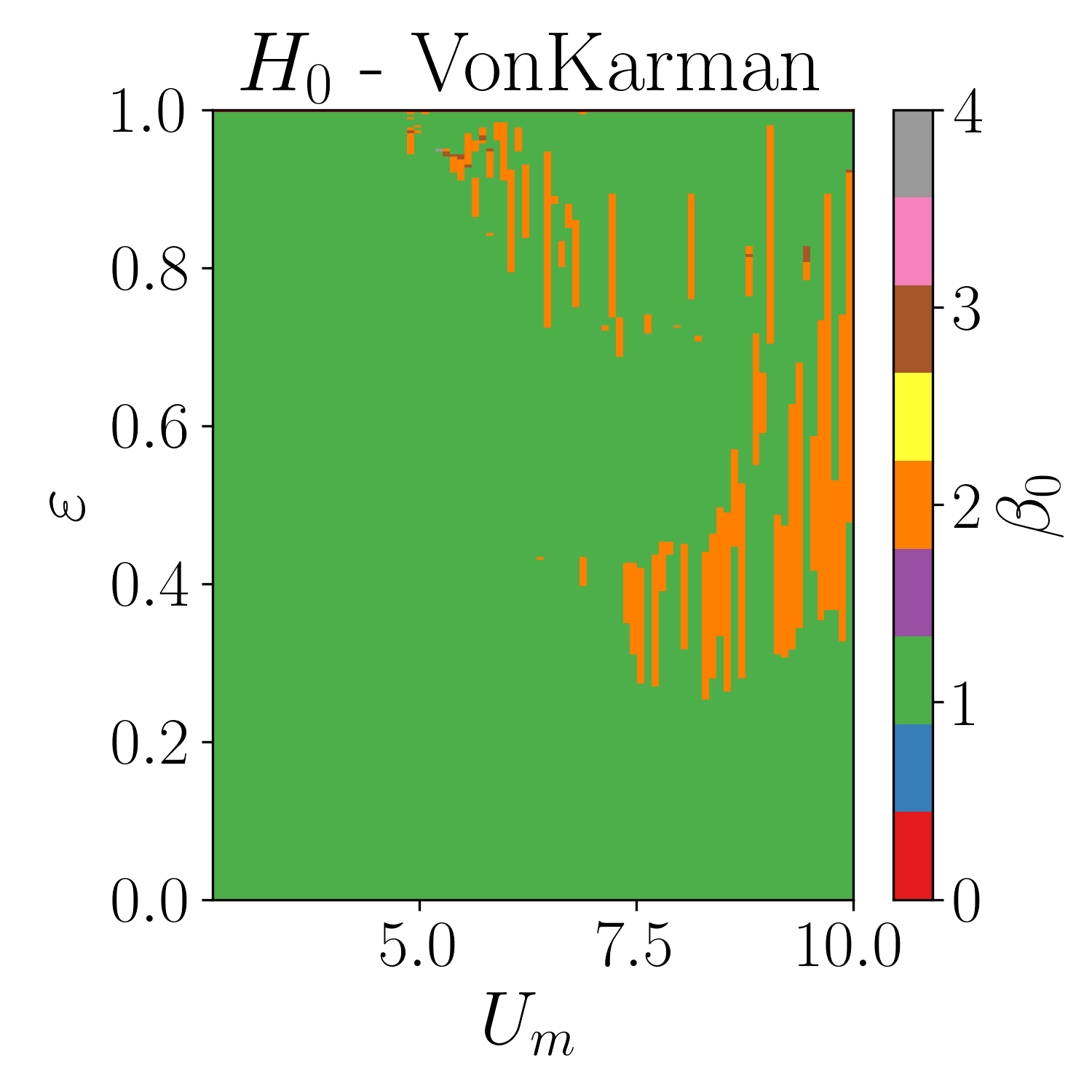}

\includegraphics[width=0.32\textwidth]{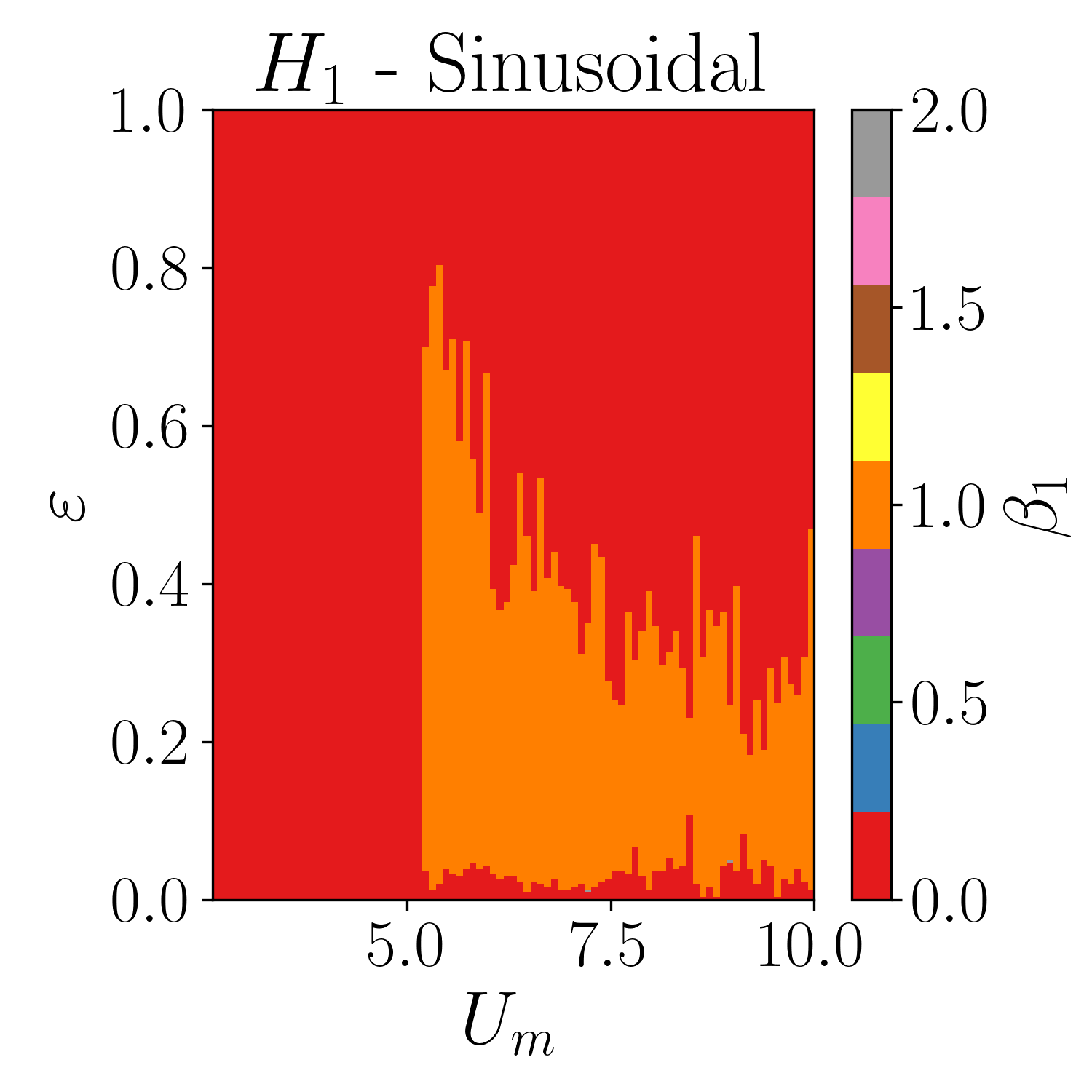}
\includegraphics[width=0.32\textwidth]{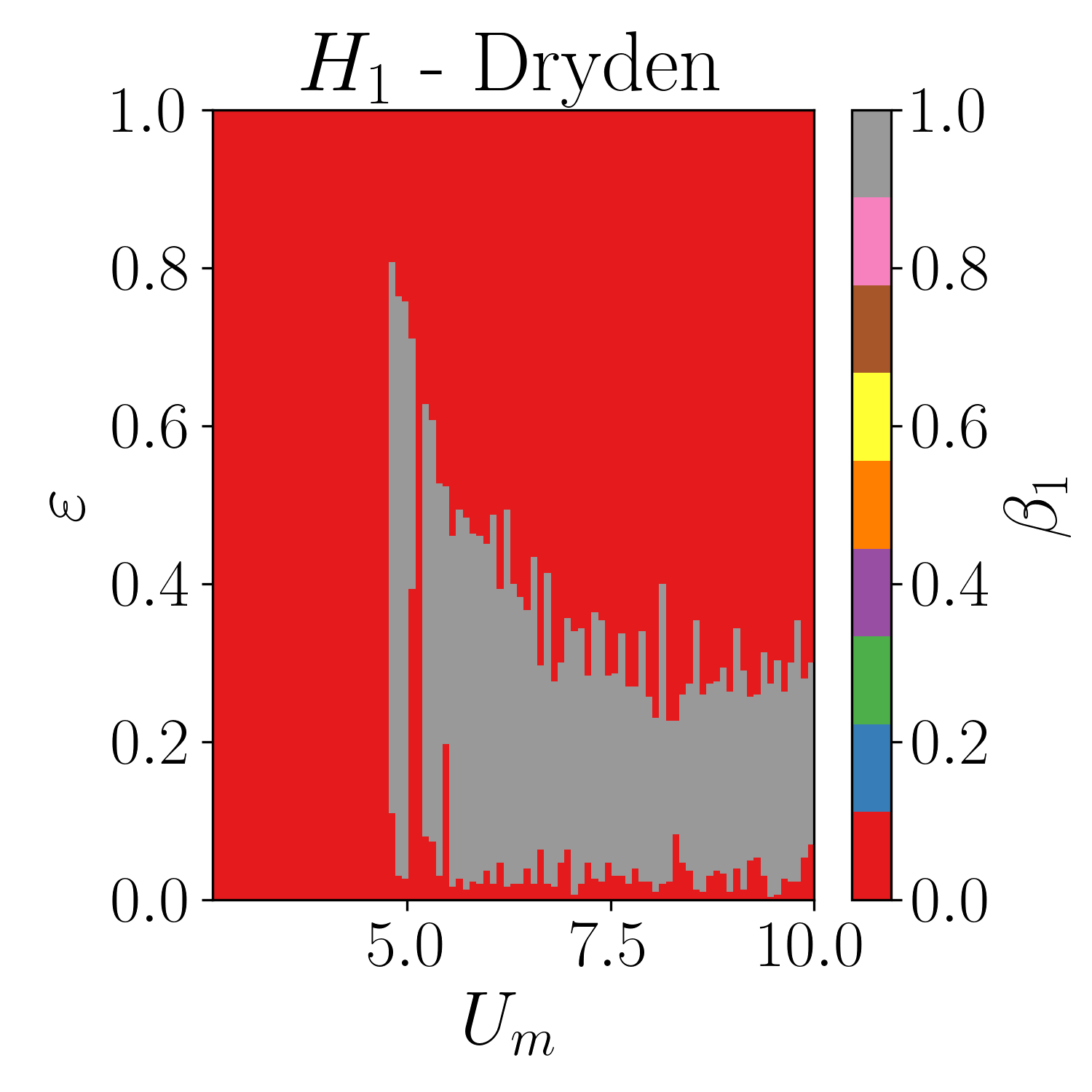}
\includegraphics[width=0.32\textwidth]{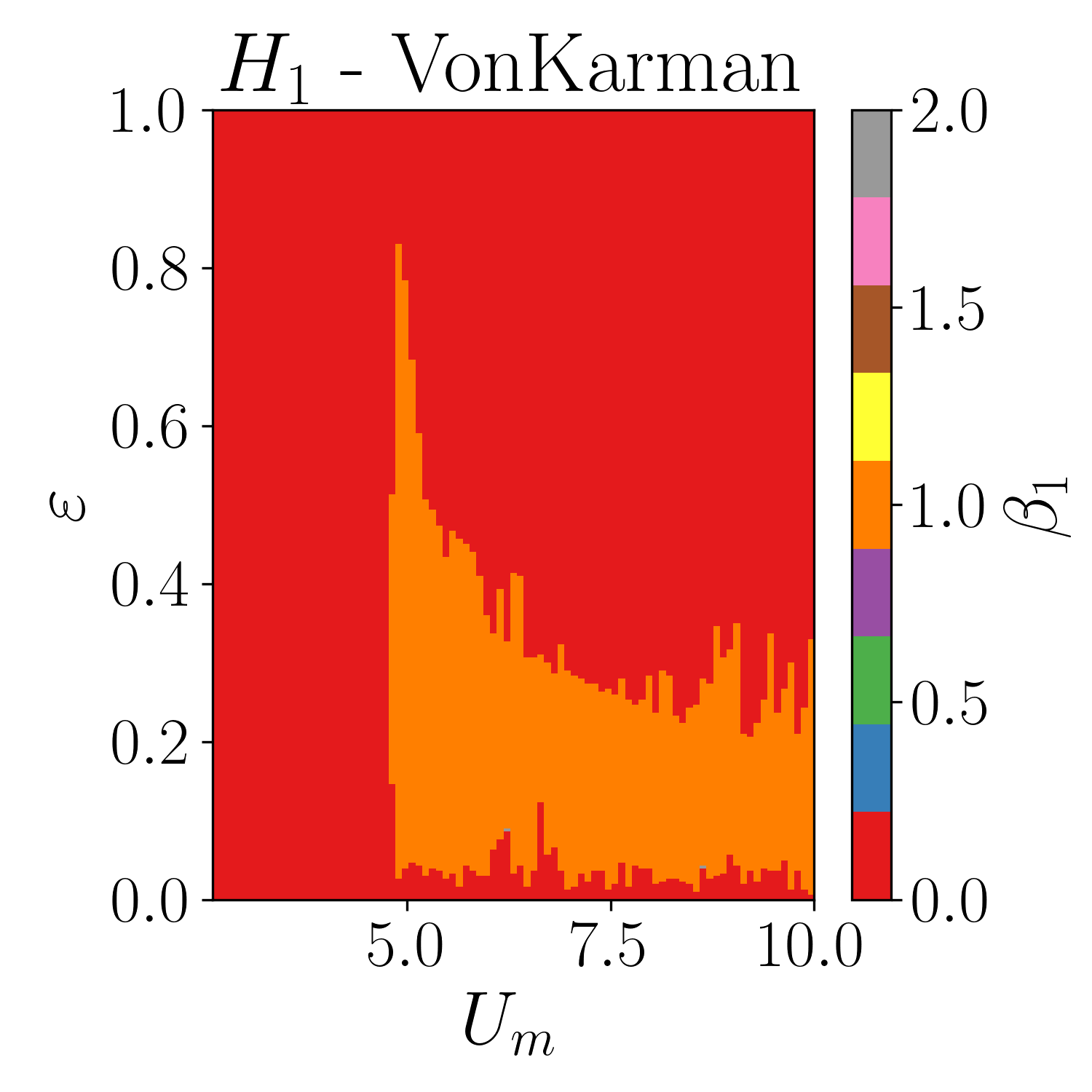}

\includegraphics[width=0.32\textwidth]{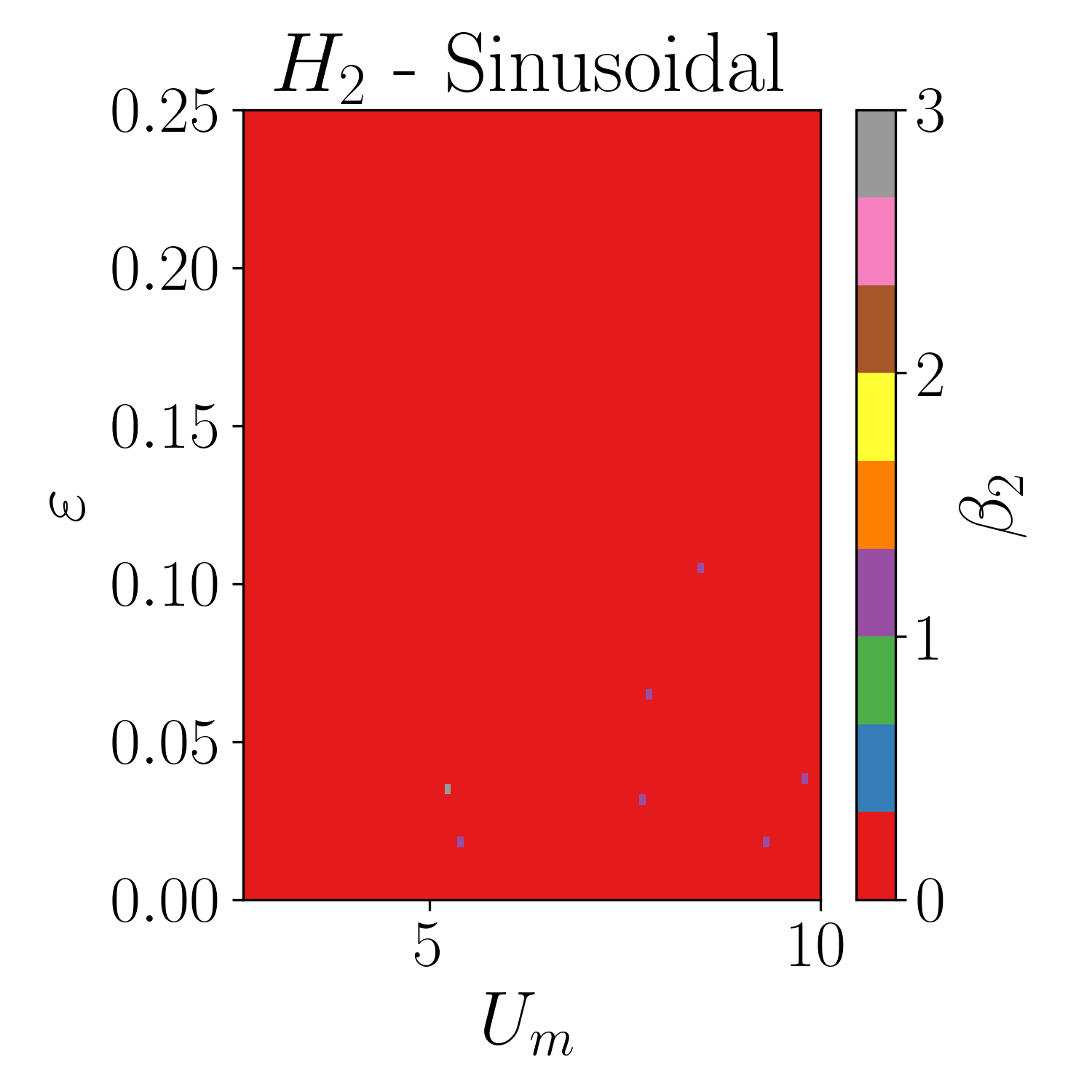}
\includegraphics[width=0.32\textwidth]{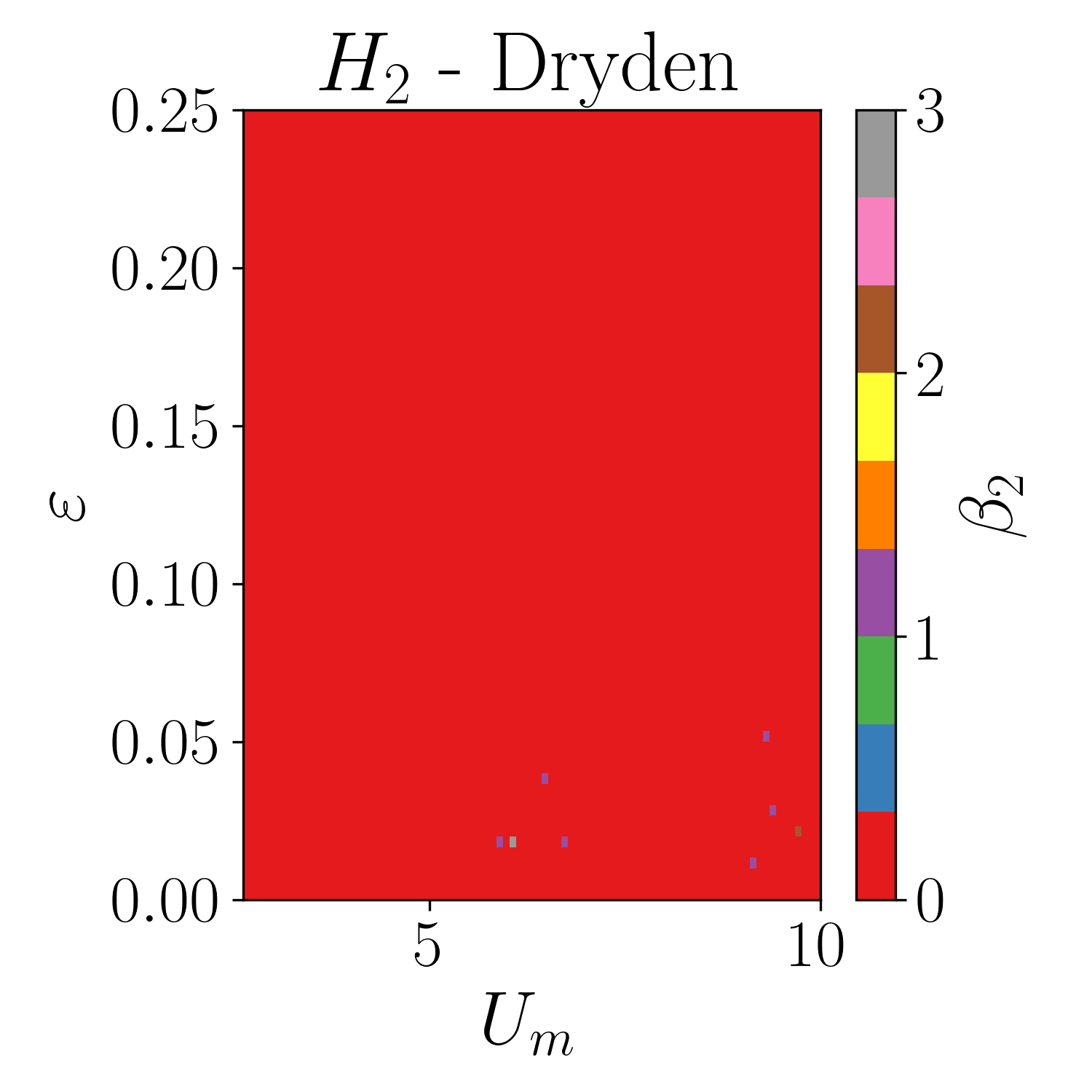}
\includegraphics[width=0.32\textwidth]{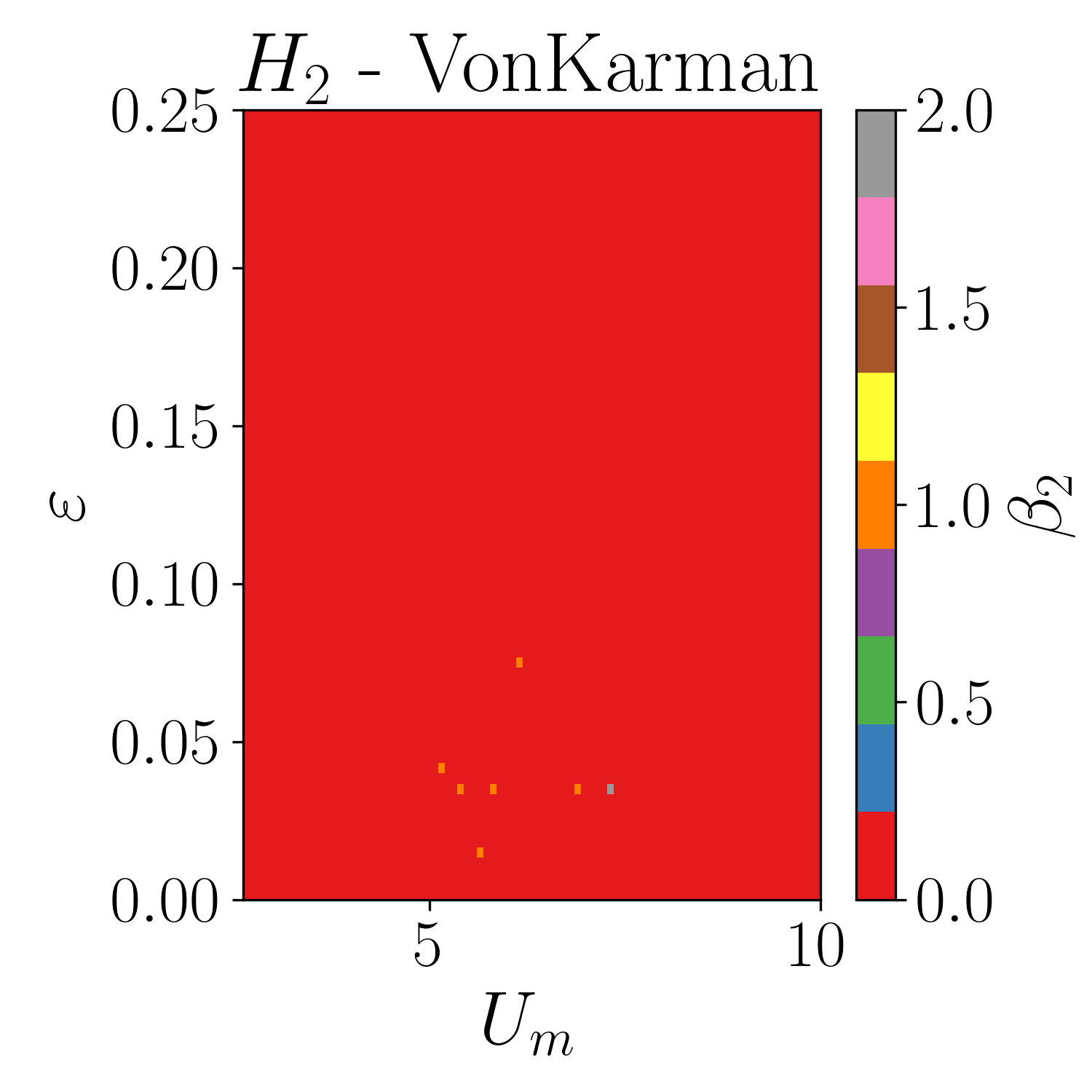}

\caption{Homological bifurcation plots for the sinusoidal, Dryden, and von Kármán models.  The x-axis is the mean flow speed \(U_m\), the y-axis is the normalized filtration level, and color represents the number of connected components \(\beta_p\). Note: The scale for $H_2$ is intentionally cropped for as the $\beta_2 > 0$ values appear for small filtrations only.}

\label{fig:homo_all}
\end{figure*}

Figure~\ref{fig:homo_all} shows the homological bifurcation plots for the three turbulence models across the first three homology dimensions.  
The horizontal axis corresponds to the mean flow speed \(U_m\), and the vertical axis corresponds 
to the superlevel parameter \(\varepsilon\) of the cubical complex filtration.
Each color indicates the Betti number \(\beta_k\) detected at that resolution, allowing us to 
identify when new connected components (\(H_0\)), loops (\(H_1\)), or voids (\(H_2\)) emerge in 
the stationary probability density. Despite only modest differences between the noise models, the homological bifurcation plots robustly detect and distinguish their behavior. 

\paragraph{Connected Components (\(H_0\)):} For sinusoidal excitation, the density remains unimodal until approximately \(U_m \approx 5.5\) m/s, at which point the homological signature transitions from 
\(\beta_0 = 1\) to \(\beta_0 = 2\). This transition is sharp and confined to a relatively narrow band of \(\varepsilon\), consistent with the delayed instability onset seen in the time-domain dynamics. In contrast, both Dryden and von Kármán turbulence exhibit earlier and less abrupt transitions in \(H_0\).  
Under correlated turbulence, the density splits into multiple component as early as \(U_m \approx 5\) m/s, and this multi-component structure persists over a broader range of superlevel thresholds. This earlier appearance of \(\beta_0 = 2\) (and occasionally \(\beta_0 = 3,4\)) reflects the fact that colored stochastic fluctuations can intermittently push the system toward flutter-like oscillations before the deterministic threshold.

\paragraph{Limit-Cycle Loops (\(H_1\)):} The transition in the first homology group occurs around the same speed as \(H_0\) transitions. For sinusoidal excitation, persistent \(H_1\) loops emerge around \(U_m \approx 5.5\) m/s, matching the onset of clean, periodic limit cycles in the phase-space trajectories. Dryden and von Kármán turbulence again produce earlier transitions with \(\beta_1 > 0\) appearing as early as \(U_m \approx 4.75\) m/s. In these models, the \(H_1\) “bifurcation tongue’’ slopes downward with increasing \(U_m\), 
indicating that while loops are born early, they persist mainly at intermediate probability levels.

\paragraph{Higher-Dimensional Features (\(H_2\)):} 
The second homology group exhibits only weak and short-lived features across all three excitation models. While isolated regions with \(\beta_2 > 0\) appear for \(U_m > 5\) m/s, these features are not persistent across filtration levels. Interestingly, low-dimensional projections of the stationary distribution (e.g., pairwise marginals) for these would exhibit annular or ``volcano-like'' structures, suggesting the presence of hollow regions. However, the lack of persistent \(H_2\) features indicates that these apparent cavities are projection artifacts arising from the stochastic thickening of limit-cycle dynamics, rather than true higher-dimensional voids.
Thus, while the density would appear shell-like in projections, its intrinsic topology is dominated by a single loop structure (\(H_1\)), consistent with a noise-broadened limit cycle rather than a toroidal manifold.

\section{Conclusion}

This work compares the effect of different noise models on appearance of stochastic bifurcations in a two-degree-of-freedom aeroelastic system subject to three forms of parametric excitation---sinusoidal fluctuations, Dryden turbulence, and von Kármán turbulence. By constructing full phase-space probability densities and analyzing their topology, we identified transitions in \(\beta_0\), \(\beta_1\), and \(\beta_2\) that signal the 
appearance of multi-modal distributions, limit-cycle loops, and voids respectively. The homological bifurcation plots revealed that each excitation model induces a distinct bifurcation pattern. Sinusoidal forcing produced the latest transitions while both Dryden and von Kármán turbulence generated earlier transitions. The correlated nature of these turbulence models accelerates the onset of flutter-like behavior. Although the second homology \(\beta_2\) appeared less frequently, isolated pockets of non-zero \(\beta_2\) were detected for all three models at flow speeds beyond approximately \(U_m \approx 5\)~m/s. Although pairwise projections of the stationary distribution would exhibit annular (``volcano-like") structures, the topological analysis shows that these do not correspond to persistent higher-dimensional voids. Instead, the dominant structure is a loop (\(\beta_1 = 1\)) arising from a stochastic thickening of limit-cycle dynamics. This highlights the importance of topological methods in distinguishing intrinsic geometry from projection-induced artifacts.

Although our analysis used an example stochastic flutter model, the approach we present can be applied to different models and with different gust conditions. It can further be automated to conduct large-scale parameter studies on flutter phenomenon in high-dimensional aeroelastic systems and may support future efforts in uncertainty quantification, reduced-order modeling, and real-time monitoring of aeroelastic behavior.

\section{Acknowledgment}
This material is based upon work supported by the Air Force Office of Scientific Research under award number FA9550-26-1-0011, resources from Texas Advanced Computing Center at University of Texas-Austin, and by the NSF Frontera Computational Science Fellowship awarded to ST for 2025-2026. 

\appendix
\section*{Topological Data Analysis}
\label{app:topological-background}

\subsection*{Homology and Betti Numbers}

Homology provides a quantitative description of the topological structure of a space by identifying its connected components, loops, and voids.  
For a topological space \(X\), the \(p\)-th homology group \(H_p(X)\) encodes \(p\)-dimensional holes, and its rank is called the \textit{Betti number}:
\[
\beta_p(X) = \dim H_p(X).
\]
Intuitively, \(\beta_0\) counts connected components, \(\beta_1\) counts one-dimensional loops, and \(\beta_2\) counts enclosed cavities or voids.

When homology is computed across a sequence of spaces related by inclusion, we obtain a \emph{filtration}:
\[
H_p(X_{a_1}) \rightarrow H_p(X_{a_2}) \rightarrow \cdots \rightarrow H_p(X_{a_n}),
\]
where \(a_i\) denotes the value of the filtration parameter.  
Persistent homology studies the appearance (birth) and disappearance (death) of topological features across this filtration.

\subsection*{Cubical Complexes and Superlevel Persistence}

In this study, the state-space kernel density estimate (KDE) is discretized into an \(m \times n \times k \times l\) array, forming a \textit{cubical complex} \(K\).  
Each cube \(s_{i,j,k,l}\) is assigned a scalar value equal to its probability density, giving a function \(f: K \to \mathbb{R}\).  
For a threshold \(L\), the corresponding \textit{superlevelset} is defined as
\[
K_L = f^{-1}([L, \infty)),
\]
which collects all cubes with density greater than or equal to \(L\).  
As \(L\) decreases from the peak of the PDF to zero, high-probability regions merge, loops form, and cavities appear or vanish, representing the geometric structure of the probability distribution.

\begin{figure}[!htbp]
    \centering
    \includegraphics[width=0.4\textwidth]{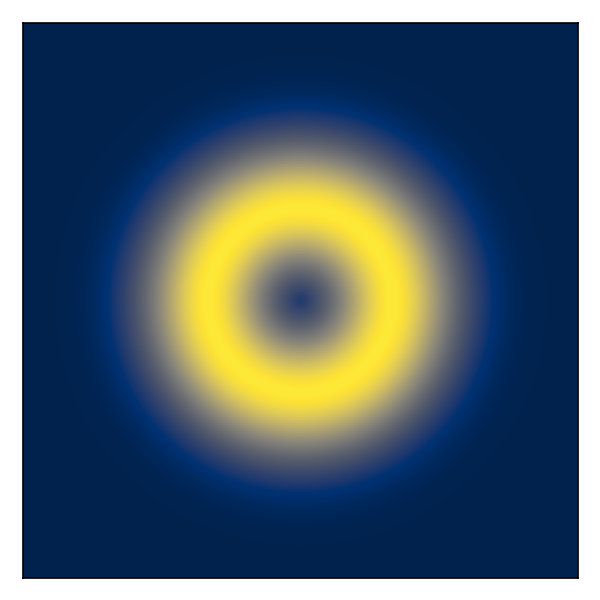}
    \includegraphics[width=0.4\linewidth]{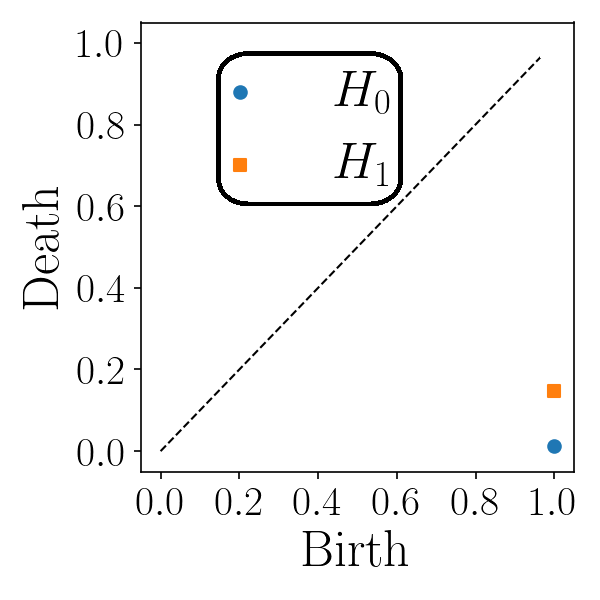}
    \includegraphics[width=0.4\linewidth]{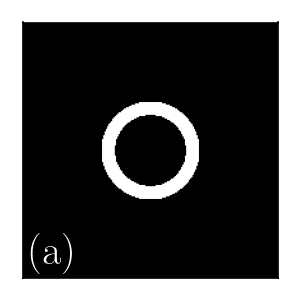}
    \includegraphics[width=0.4\linewidth]{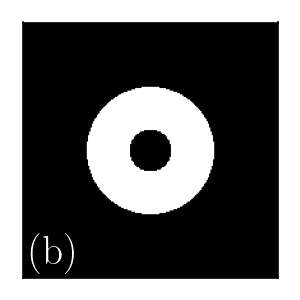}
    \includegraphics[width=0.4\linewidth]{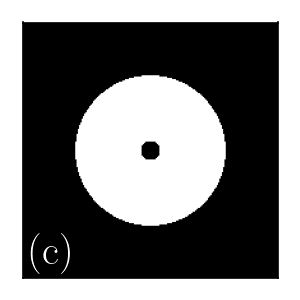}
    \includegraphics[width=0.4\linewidth]{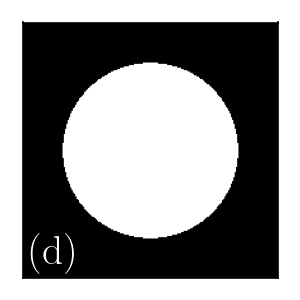}
    \caption{Illustration of superlevel-set persistence for an annular probability density. Panels (a-d) show the binary superlevel sets of the field at thresholds 
    $L = 0.95, 0.6, 0.3$ and $0.1$, respectively. 
    At high thresholds, only the most intense region of the ring remains visible, corresponding to a single connected component and loop. As the filtration level decreases, the ring thickens and closes the loop.}
    \label{fig:appendix}
\end{figure}

This family of nested complexes
\[
K_{L_1} \subseteq K_{L_2} \subseteq \cdots \subseteq K_{L_N}
\]
defines a \emph{superlevel filtration}.  
The homology of each \(K_L\) is computed to yield the Betti numbers
\(\beta_p(L)\), and their variation over \(L\) defines a \textit{Betti curve}.  
In discretized settings, the collection of values
\[
[\beta_p(L_0), \beta_p(L_1), \ldots, \beta_p(L_N)]
\]
is referred to as a \textit{Betti vector}.  
In this context, the filtration parameter \(L\) corresponds directly to the normalized probability density of the aeroelastic system’s stationary KDE. See Fig.~\ref{fig:appendix} for an illustrative example.

\subsection*{Persistent Homology and Bifurcation Analysis}

By repeating the above computation across a range of bifurcation parameters---here, the mean flow speed \(U_m\)---one obtains a two-dimensional field of Betti numbers:
\[
\beta_p(U_m, L),
\]
which is visualized as a \emph{homological bifurcation plot}.  
Color represents the magnitude of the Betti number, the \(x\)-axis corresponds to the bifurcation parameter \(U_m\), and the \(y\)-axis corresponds to the filtration threshold \(L\).  
Abrupt changes in \(\beta_0\), \(\beta_1\), or \(\beta_2\) signify topological transitions in the system’s invariant measure. 

\section{Topology-Informed Kernel Density Estimation}
\label{app:kde_unsupervised}

We use the method in~\cite{UnsupervisedKDEPaper} to estimate the KDE in high dimensions. The method is summarised here.

\subsection*{Kernel Density Estimation and Bandwidth Parameter}
Given samples $\{x_i\}_{i=1}^n \subset \mathbb{R}^d$, the Gaussian kernel density
estimate with bandwidth $h>0$ is
\[
\widehat{f}_h(x)
=
\frac{1}{n h^d}
\sum_{i=1}^n 
\exp\!\left(-\frac{\|x - x_i\|^2}{2 h^2}\right).
\]
The bandwidth $h$ governs the smoothness of the estimate: small $h$ produces
undersmoothing (spurious oscillations), while large $h$ causes oversmoothing and
loss of structure.  Our objective is to select $h$ through a purely unsupervised,
topology-driven loss.

\subsection*{Persistence Diagram and Topological Summary}
For each bandwidth $h$, computing superlevel persistence yields a persistence diagram $D_h$ containing all homological features across all thresholds $L$.
For the purpose of bandwidth selection, we restrict attention to $H_0$ (connected
components) and measure:
\begin{itemize}
    \item the \emph{betti number} 
    \[
    \mathrm{count}(h) = |D_h|,
    \]
    \item the \emph{total persistence}
    \[
    \mathrm{TP}(h) = \sum_{(b,d)\in D_h} (b-d).
    \]
\end{itemize}
These quantities summarize how structured or oscillatory the KDE is at bandwidth $h$.

\subsection*{Topology-Based Loss Function}
The bandwidth is chosen by minimizing a loss composed only of these two terms:
\[
\mathcal{L}(h)
  = \alpha_{\mathrm{TP}} \, \mathrm{TP}(h)
  + \alpha_{\mathrm{count}} \, \mathrm{count}(h),
\]
where $\alpha_{\mathrm{TP}}$ and $\alpha_{\mathrm{count}}$ are fixed nonnegative
weights (typically both set to~1 in our experiments).  
Larger bandwidths reduce both terms but oversmooth the density, eliminating meaningful peaks; smaller bandwidths inflate both terms by creating many spurious components. Thus, the loss is minimized at an intermediate, data-driven scale. Since $\mathcal{L}(h)$ is one-dimensional, we optimize $h$ using stochastic gradient descent. The gradient is obtained through differentiable KDE combined with a differentiable persistence pipeline.

\bibliographystyle{AAS_publication}
\bibliography{SDE}

\begin{thebibliography}{10}

\bibitem{Berci2021}
M.~Berci, ``On Aerodynamic Models for Flutter Analysis: A Systematic Overview and Comparative Assessment,''  {\em Applied Mechanics}, Vol.~2, July 2021, p.~516–541, 10.3390/applmech2030029.

\bibitem{Jones1996}
K.~D. Jones and M.~F. Platzer, ``Time-domain analysis of low-speed airfoil flutter,''  {\em AIAA Journal}, Vol.~34, May 1996, p.~1027–1033, 10.2514/3.13183.

\bibitem{Venkatramani2018}
J.~Venkatramani, S.~Sarkar, and S.~Gupta, ``Intermittency in pitch-plunge aeroelastic systems explained through stochastic bifurcations,''  {\em Nonlinear Dynamics}, Vol.~92, Feb. 2018, p.~1225–1241, 10.1007/s11071-018-4121-5.

\bibitem{Chen2022}
Z.~Chen, Z.~Shi, S.~Chen, and Z.~Yao, ``Stall flutter suppression of NACA 0012 airfoil based on steady blowing,''  {\em Journal of Fluids and Structures}, Vol.~109, Feb. 2022, p.~103472, 10.1016/j.jfluidstructs.2021.103472.

\bibitem{Huang1987}
X.~Y. Huang, ``Active control of aerofoil flutter,''  {\em AIAA Journal}, Vol.~25, Aug. 1987, p.~1126–1132, 10.2514/3.9753.

\bibitem{Wu2022}
Y.~Wu, Y.~Dai, and C.~Yang, ``Time-Delayed Active Control of Stall Flutter for an Airfoil via Camber Morphing,''  {\em AIAA Journal}, Vol.~60, Oct. 2022, p.~5723–5734, 10.2514/1.j061947.

\bibitem{SriNamachchivaya1990}
N.~Sri~Namachchivaya, ``Stochastic bifurcation,''  {\em Applied Mathematics and Computation}, Vol.~38, July 1990, p.~101–159, 10.1016/0096-3003(90)90051-4.

\bibitem{Bass1999}
R.~F. Bass and K.~Burdzy, ``Stochastic Bifurcation Models,''  {\em The Annals of Probability}, Vol.~27, Jan. 1999, 10.1214/aop/1022677254.

\bibitem{Ketseas2024}
D.~Ketseas, ``Stochastic Response of an Airfoil and Its Effects on LCO’s Behavior Under Stall Flutter Regime,''  {\em International Journal of Mathematics, Statistics, and Computer Science}, Vol.~2, Jan. 2024, p.~168–172, 10.59543/ijmscs.v2i.8663.

\bibitem{Bethi2018}
R.~V. Bethi, S.~V. Reddy~Gali, and J.~Venkatramani, ``Identifying route to stall flutter through stochastic bifurcation analysis,''  {\em MATEC Web of Conferences}, Vol.~211, 2018, p.~02011, 10.1051/matecconf/201821102011.

\bibitem{Irani2016}
S.~Irani, S.~Sazesh, and V.~R. Molazadeh, ``Flutter analysis of a nonlinear airfoil using stochastic approach,''  {\em Nonlinear Dynamics}, Vol.~84, Jan. 2016, p.~1735–1746, 10.1007/s11071-016-2601-z.

\bibitem{Hao2018}
Y.~Hao and Z.~Q. Wu, ``Random Flutter of Multi-Stable Airfoils Excited Parametrically in Steady Flows,''  {\em Journal of Mechanics}, Vol.~35, July 2018, p.~419–426, 10.1017/jmech.2018.19.

\bibitem{Tripathi2022}
D.~Tripathi, R.~Shreenivas, C.~Bose, S.~Mondal, and J.~Venkatramani, ``Experimental investigation on the synchronization characteristics of a pitch-plunge aeroelastic system exhibiting stall flutter,''  {\em Chaos: An Interdisciplinary Journal of Nonlinear Science}, Vol.~32, July 2022, 10.1063/5.0096213.

\bibitem{Beal1993}
T.~R. Beal, ``Digital simulation of atmospheric turbulence for Dryden and von Karman models,''  {\em Journal of Guidance, Control, and Dynamics}, Vol.~16, Jan. 1993, p.~132–138, 10.2514/3.11437.

\bibitem{Arnold1998}
L.~Arnold, {\em Random Dynamical Systems}.
\newblock Springer Berlin Heidelberg, 1998, 10.1007/978-3-662-12878-7.

\bibitem{Crauel1994}
H.~Crauel and F.~Flandoli, ``Attractors for random dynamical systems,''  {\em Probability Theory and Related Fields}, Vol.~100, Sept. 1994, p.~365–393, 10.1007/bf01193705.

\bibitem{Tanweer2024}
S.~Tanweer, F.~A.~Khasawneh, E.~Munch, and J.~R.~Tempelman, ``A topological framework for identifying phenomenological bifurcations in stochastic dynamical systems,''  {\em Nonlinear Dynamics}, Vol.~112, Feb. 2024, p.~4687–4703, 10.1007/s11071-024-09289-1.

\bibitem{UnsupervisedKDEPaper}
S.~Tanweer and F.~A. Khasawneh, ``Unsupervised Learning of Density Estimates with Topological Optimization,''  2025, 10.48550/ARXIV.2512.08895.

\end{thebibliography}

\end{document}